\documentclass{wagasci_paper}

\usepackage{amsmath} %for dealing with mathematics,
\usepackage{amsthm} %for dealing with theorem environments,
\usepackage{algorithmic} %for describing algorithms
\usepackage{subfig} %for getting the subfigures e.g., "Figure 1a and 1b" etc.
\usepackage{comment}
\usepackage{multirow}
\usepackage{here}
\usepackage{url}
\def\WP{the WAGASCI module and the Proton Module}
\def\nmb{$\overline{\nu}_\mu$~}
\def\nm{$\nu_\mu$~}
\def\nmbm{$\overline{\nu}_\mu+\nu_\mu$~}
\def\neb{$\overline{\nu}_e$}
\def\ne{$\nu_e$}

\def\hto{$\rm H_2O$}
\def\sigho{$\rm \sigma_{H_2O}$}
\def\sigch{$\rm \sigma_{CH}$}
\def\sigratio{$\rm \sigma_{H_2O}/\sigma_{CH}$}

\def\ccnopp{$\rm CC0\pi0p$}
\def\ccnoppf{$\rm CC0\pi0p$ with a muon angle smaller than 30\,degrees}

\newcommand*{\tb}[1]{C_{#1}^{\rm true}}
\newcommand*{\rb}[1]{X_{#1}^{\rm reco}}

\begin{document}

\title{Measurements of $\overline{\nu}_{\mu}$ and $\overline{\nu}_{\mu} + \nu_{\mu}$ charged-current cross-sections without detected pions nor protons on water and hydrocarbon at mean antineutrino energy of 0.86 GeV}

%%%% To generate auto affiliation numbers please use \author{}\affil{} command
\setcounter{footnote}{0}
\newcommand{\INSTHD}{\affiliation[1]{University Autonoma Madrid, Department of Theoretical Physics, 28049 Madrid, Spain}}
\newcommand{\INSTEE}{\affiliation[2]{University of Bern, Albert Einstein Center for Fundamental Physics, Laboratory for High Energy Physics (LHEP), Bern, Switzerland}}
\newcommand{\INSTFE}{\affiliation[3]{Boston University, Department of Physics, Boston, Massachusetts, U.S.A.}}
\newcommand{\INSTD}{\affiliation[4]{University of British Columbia, Department of Physics and Astronomy, Vancouver, British Columbia, Canada}}
\newcommand{\INSTGA}{\affiliation[5]{University of California, Irvine, Department of Physics and Astronomy, Irvine, California, U.S.A.}}
\newcommand{\INSTI}{\affiliation[6]{IRFU, CEA Saclay, Gif-sur-Yvette, France}}
\newcommand{\INSTGB}{\affiliation[7]{University of Colorado at Boulder, Department of Physics, Boulder, Colorado, U.S.A.}}
\newcommand{\INSTFG}{\affiliation[8]{Colorado State University, Department of Physics, Fort Collins, Colorado, U.S.A.}}
\newcommand{\INSTFH}{\affiliation[9]{Duke University, Department of Physics, Durham, North Carolina, U.S.A.}}
\newcommand{\INSTBA}{\affiliation[10]{Ecole Polytechnique, IN2P3-CNRS, Laboratoire Leprince-Ringuet, Palaiseau, France }}
\newcommand{\INSTEF}{\affiliation[11]{ETH Zurich, Institute for Particle Physics and Astrophysics, Zurich, Switzerland}}
\newcommand{\INSTIE}{\affiliation[12]{CERN European Organization for Nuclear Research, CH-1211 Genève 23, Switzerland}}
\newcommand{\INSTEG}{\affiliation[13]{University of Geneva, Section de Physique, DPNC, Geneva, Switzerland}}
\newcommand{\INSTHJ}{\affiliation[14]{University of Glasgow, School of Physics and Astronomy, Glasgow, United Kingdom}}
\newcommand{\INSTDG}{\affiliation[15]{H. Niewodniczanski Institute of Nuclear Physics PAN, Cracow, Poland}}
\newcommand{\INSTCB}{\affiliation[16]{High Energy Accelerator Research Organization (KEK), Tsukuba, Ibaraki, Japan}}
\newcommand{\INSTIB}{\affiliation[17]{University of Houston, Department of Physics, Houston, Texas, U.S.A.}}
\newcommand{\INSTED}{\affiliation[18]{Institut de Fisica d'Altes Energies (IFAE), The Barcelona Institute of Science and Technology, Campus UAB, Bellaterra (Barcelona) Spain}}
\newcommand{\INSTEC}{\affiliation[19]{IFIC (CSIC \& University of Valencia), Valencia, Spain}}
\newcommand{\INSTHH}{\affiliation[20]{Institute For Interdisciplinary Research in Science and Education (IFIRSE), ICISE, Quy Nhon, Vietnam}}
\newcommand{\INSTEI}{\affiliation[21]{Imperial College London, Department of Physics, London, United Kingdom}}
\newcommand{\INSTGF}{\affiliation[22]{INFN Sezione di Bari and Universit\`a e Politecnico di Bari, Dipartimento Interuniversitario di Fisica, Bari, Italy}}
\newcommand{\INSTBE}{\affiliation[23]{INFN Sezione di Napoli and Universit\`a di Napoli, Dipartimento di Fisica, Napoli, Italy}}
\newcommand{\INSTBF}{\affiliation[24]{INFN Sezione di Padova and Universit\`a di Padova, Dipartimento di Fisica, Padova, Italy}}
\newcommand{\INSTBD}{\affiliation[25]{INFN Sezione di Roma and Universit\`a di Roma ``La Sapienza'', Roma, Italy}}
\newcommand{\INSTEB}{\affiliation[26]{Institute for Nuclear Research of the Russian Academy of Sciences, Moscow, Russia}}
\newcommand{\INSTHI}{\affiliation[27]{International Centre of Physics, Institute of Physics (IOP), Vietnam Academy of Science and Technology (VAST), 10 Dao Tan, Ba Dinh, Hanoi, Vietnam}}
\newcommand{\INSTHA}{\affiliation[28]{Kavli Institute for the Physics and Mathematics of the Universe (WPI), The University of Tokyo Institutes for Advanced Study, University of Tokyo, Kashiwa, Chiba, Japan}}
\newcommand{\INSTID}{\affiliation[29]{Keio University, Department of Physics, Kanagawa, Japan}}
\newcommand{\INSTIF}{\affiliation[30]{King's College London, Department of Physics, Strand, London WC2R 2LS, United Kingdom}}
\newcommand{\INSTCC}{\affiliation[31]{Kobe University, Kobe, Japan}}
\newcommand{\INSTCD}{\affiliation[32]{Kyoto University, Department of Physics, Kyoto, Japan}}
\newcommand{\INSTEJ}{\affiliation[33]{Lancaster University, Physics Department, Lancaster, United Kingdom}}
\newcommand{\INSTFC}{\affiliation[34]{University of Liverpool, Department of Physics, Liverpool, United Kingdom}}
\newcommand{\INSTFI}{\affiliation[35]{Louisiana State University, Department of Physics and Astronomy, Baton Rouge, Louisiana, U.S.A.}}
\newcommand{\INSTHB}{\affiliation[36]{Michigan State University, Department of Physics and Astronomy,  East Lansing, Michigan, U.S.A.}}
\newcommand{\INSTCE}{\affiliation[37]{Miyagi University of Education, Department of Physics, Sendai, Japan}}
\newcommand{\INSTDF}{\affiliation[38]{National Centre for Nuclear Research, Warsaw, Poland}}
\newcommand{\INSTFJ}{\affiliation[39]{State University of New York at Stony Brook, Department of Physics and Astronomy, Stony Brook, New York, U.S.A.}}
\newcommand{\INSTGJ}{\affiliation[40]{Okayama University, Department of Physics, Okayama, Japan}}
\newcommand{\INSTCF}{\affiliation[41]{Osaka City University, Department of Physics, Osaka, Japan}}
\newcommand{\INSTGG}{\affiliation[42]{Oxford University, Department of Physics, Oxford, United Kingdom}}
\newcommand{\INSTIC}{\affiliation[43]{University of Pennsylvania, Department of Physics and Astronomy,  Philadelphia, PA, 19104, USA.}}
\newcommand{\INSTGC}{\affiliation[44]{University of Pittsburgh, Department of Physics and Astronomy, Pittsburgh, Pennsylvania, U.S.A.}}
\newcommand{\INSTFA}{\affiliation[45]{Queen Mary University of London, School of Physics and Astronomy, London, United Kingdom}}
\newcommand{\INSTE}{\affiliation[46]{University of Regina, Department of Physics, Regina, Saskatchewan, Canada}}
\newcommand{\INSTGD}{\affiliation[47]{University of Rochester, Department of Physics and Astronomy, Rochester, New York, U.S.A.}}
\newcommand{\INSTHC}{\affiliation[48]{Royal Holloway University of London, Department of Physics, Egham, Surrey, United Kingdom}}
\newcommand{\INSTBC}{\affiliation[49]{RWTH Aachen University, III. Physikalisches Institut, Aachen, Germany}}
\newcommand{\INSTFB}{\affiliation[50]{University of Sheffield, Department of Physics and Astronomy, Sheffield, United Kingdom}}
\newcommand{\INSTDI}{\affiliation[51]{University of Silesia, Institute of Physics, Katowice, Poland}}
\newcommand{\INSTIA}{\affiliation[52]{SLAC National Accelerator Laboratory, Stanford University, Menlo Park, California, USA}}
\newcommand{\INSTBB}{\affiliation[53]{Sorbonne Universit\'e, Universit\'e Paris Diderot, CNRS/IN2P3, Laboratoire de Physique Nucl\'eaire et de Hautes Energies (LPNHE), Paris, France}}
\newcommand{\INSTEH}{\affiliation[54]{STFC, Rutherford Appleton Laboratory, Harwell Oxford,  and  Daresbury Laboratory, Warrington, United Kingdom}}
\newcommand{\INSTCH}{\affiliation[55]{University of Tokyo, Department of Physics, Tokyo, Japan}}
\newcommand{\INSTBJ}{\affiliation[56]{University of Tokyo, Institute for Cosmic Ray Research, Kamioka Observatory, Kamioka, Japan}}
\newcommand{\INSTCG}{\affiliation[57]{University of Tokyo, Institute for Cosmic Ray Research, Research Center for Cosmic Neutrinos, Kashiwa, Japan}}
\newcommand{\INSTHF}{\affiliation[58]{Tokyo Institute of Technology, Department of Physics, Tokyo, Japan}}
\newcommand{\INSTGI}{\affiliation[59]{Tokyo Metropolitan University, Department of Physics, Tokyo, Japan}}
\newcommand{\INSTHG}{\affiliation[60]{Tokyo University of Science, Faculty of Science and Technology, Department of Physics, Noda, Chiba, Japan}}
\newcommand{\INSTF}{\affiliation[61]{University of Toronto, Department of Physics, Toronto, Ontario, Canada}}
\newcommand{\INSTB}{\affiliation[62]{TRIUMF, Vancouver, British Columbia, Canada}}
\newcommand{\INSTG}{\affiliation[63]{University of Victoria, Department of Physics and Astronomy, Victoria, British Columbia, Canada}}
\newcommand{\INSTDJ}{\affiliation[64]{University of Warsaw, Faculty of Physics, Warsaw, Poland}}
\newcommand{\INSTDH}{\affiliation[65]{Warsaw University of Technology, Institute of Radioelectronics and Multimedia Technology, Warsaw, Poland}}
\newcommand{\INSTFD}{\affiliation[66]{University of Warwick, Department of Physics, Coventry, United Kingdom}}
\newcommand{\INSTGH}{\affiliation[67]{University of Winnipeg, Department of Physics, Winnipeg, Manitoba, Canada}}
\newcommand{\INSTEA}{\affiliation[68]{Wroclaw University, Faculty of Physics and Astronomy, Wroclaw, Poland}}
\newcommand{\INSTHE}{\affiliation[69]{Yokohama National University, Faculty of Engineering, Yokohama, Japan}}
\newcommand{\INSTH}{\affiliation[70]{York University, Department of Physics and Astronomy, Toronto, Ontario, Canada}}

\author[56]{K.\,Abe}
\author[45]{N.\,Akhlaq}
\author[57]{R.\,Akutsu}
\author[32]{A.\,Ali}
\author[11]{C.\,Alt}
\author[54, 34]{C.\,Andreopoulos}
\author[21]{L.\,Anthony}
\author[19]{M.\,Antonova}
\author[31]{S.\,Aoki}
\author[2]{A.\,Ariga}
% ...

\INSTHD
\INSTEE
\INSTFE
\INSTD
\INSTGA
\INSTI
\INSTGB
\INSTFG
\INSTFH
\INSTBA
\INSTEF
\INSTIE
\INSTEG
\INSTHJ
\INSTDG
\INSTCB
\INSTIB
\INSTED
\INSTEC
\INSTHH
\INSTEI
\INSTGF
\INSTBE
\INSTBF
\INSTBD
\INSTEB
\INSTHI
\INSTHA
\INSTID
\INSTIF
\INSTCC
\INSTCD
\INSTEJ
\INSTFC
\INSTFI
\INSTHB
\INSTCE
\INSTDF
\INSTFJ
\INSTGJ
\INSTCF
\INSTGG
\INSTIC
\INSTGC
\INSTFA
\INSTE
\INSTGD
\INSTHC
\INSTBC
\INSTFB
\INSTDI
\INSTIA
\INSTBB
\INSTEH
\INSTCH
\INSTBJ
\INSTCG
\INSTHF
\INSTGI
\INSTHG
\INSTF
\INSTB
\INSTG
\INSTDJ
\INSTDH
\INSTFD
\INSTGH
\INSTEA
\INSTHE
\INSTH

%%% To include the collaborator name... Please use the command "\collaborator"
%%% For example: \collaborator{ATLAS Collaboration}

\begin{abstract}
We report measurements of the flux-integrated \nmb and \nmbm charged-current cross-sections on water and hydrocarbon targets using the T2K anti-neutrino beam, with a mean neutrino energy of 0.86~GeV. The signal is defined as the (anti-) neutrino charged-current interaction with one induced $\mu^\pm$ and no detected charged pion nor proton. These measurements are performed using a new WAGASCI module recently added to the T2K setup in combination with the INGRID Proton module. The phase space of muons is restricted to the high-detection efficiency region, $p_{\mu}>400~{\rm MeV}/c$ and $\theta_{\mu}<30^{\circ}$, in the laboratory frame. Absence of pions and protons in the detectable phase space of "$p_{\pi}>200~{\rm MeV}/c$ and $\theta_{\pi}<70^{\circ}$", and "$p_{\rm p}>600~{\rm MeV}/c$ and $\theta_{\rm p}<70^{\circ}$" is required. In this paper, both of the \nmb cross-sections and \nmbm cross-sections on water and hydrocarbon targets, and their ratios are provided by using D'Agostini unfolding method. The results of the integrated \nmb cross-section measurements over this phase space are $\sigma_{\rm H_{2}O}\,=\,(1.082\pm0.068(\rm stat.)^{+0.145}_{-0.128}(\rm syst.)) \times 10^{-39}~{\rm cm^{2} / nucleon}$, $\sigma_{\rm CH}\,=\,(1.096\pm0.054(\rm stat.)^{+0.132}_{-0.117}(\rm syst.)) \times 10^{-39}~{\rm cm^{2} / nucleon}$, and $\sigma_{\rm H_{2}O}/\sigma_{\rm CH} = 0.987\pm0.078(\rm stat.)^{+0.093}_{-0.090}(\rm syst.)$. The \nmbm cross-section is $\sigma_{\rm H_{2}O} = (1.155\pm0.064(\rm stat.)^{+0.148}_{-0.129}(\rm syst.)) \times 10^{-39}~{\rm cm^{2} / nucleon}$, $\sigma_{\rm CH}\,=\,(1.159\pm0.049(\rm stat.)\\^{+0.129}_{-0.115}(\rm syst.)) \times 10^{-39}~{\rm cm^{2} / nucleon}$, and $\sigma_{\rm H_{2}O}/\sigma_{\rm CH}\,=\,0.996\pm0.069(\rm stat.)^{+0.083}_{-0.078}(\rm syst.)$.
\end{abstract}

%\subjectindex{xxxx, xxx}

\maketitle

% SECTION: INTRODUCTION %%%
\section{Introduction}
The Tokai-to-Kamioka (T2K) experiment~\cite{cite:t2k} is a long baseline neutrino oscillation experiment in Japan. Using either the \nm or the \nmb beam produced at the J-PARC accelerator complex, both electron (anti-)neutrino appearance and muon (anti-)neutrino disappearance are measured at the far-detector, Super-Kamiokande (SK). T2K aims to make precision measurements of neutrino oscillation parameters, including a search for $CP$ violation in the leptonic sector by precisely measuring the (anti-)neutrino oscillation. In these measurements, the neutrino event rate at SK is constrained by the cross-section and neutrino flux measured in the near-detector, ND280. The ND280 includes two Fine-Grained Detectors, FGD1 and FGD2~\cite{cite:fgd}, used as a target for neutrino interactions and as a tracking device. The FGD1 interaction target is made up of plastic scintillators, and FGD2 consists of water and plastic scintillator targets, while SK is a water-target detector. Uncertainties in the modeling of neutrino-nucleus interactions due to the difference in the target at the near and the far-detector constitute an additional source of systematic uncertainties in the T2K oscillation analysis. In addition, unknown nuclear effects like the so called 2-particle-2-hole (2p2h) process with large uncertainties motivate testing the interaction model at multiple neutrino energies. The neutrino interaction model is used to extrapolate the neutrino beam energy distributions and interactions at the near-detector to the far-detector. Indeed, the T2K off-axis near-detector angular acceptance is more limited than the far-detector. Moreover, the near-detector event rate also includes significant interactions on materials other than the far-detector target. The interaction model is tuned from the near detector measurement and its parameterization can be incomplete. Therefore, testing the interaction model with different target materials and at various ranges of the neutrino energies are essential to improve the T2K oscillation analysis.\\
In the T2K experiment, the neutrino beam is directed 2.5 degrees off-axis with respect to the SK direction to ensure that the detector sees a narrow-band neutrino beam with a peak energy at 0.6 GeV, which maximizes the oscillation probability. In this energy range, neutrino interactions with nucleons are dominated by charged-current quasi-elastic (CCQE) and charged-current resonant-pion production (CC-Resonant). The neutrino energies from incoming CCQE interactions are reconstructed from the outgoing charged lepton kinematics. However, if multi-nucleon interactions or pion absorption occur in the nucleus, 2p2h and CC-resonant interactions may be misidentified as CCQE interactions because only a muon-like track may be observed in the final state. Furthermore, the reconstructed neutrino energy spectrum could be distorted. For this reason, in modern experiments, signals are classified by final-state particles, such as protons and pions. For example, CC0$\pi$ (charged-current interactions with no pions in the final state) cross-sections are measured instead of measuring CCQE cross-sections making them less dependent on nuclear models.

%% Motivation
So far, T2K has published two results of neutrino cross-sections on water at a mean neutrino energy of 0.6\,GeV: CC-resonant $\pi^{+}$ production cross-section using FGD2~\cite{cite:fgd_measurement} and CC0$\pi$ cross-section using a dedicated water target in the ND280 detector, called the P{\O}D~\cite{cite:pod_measurement}. CC-inclusive neutrino cross-sections using the INGRID Water Module, which consists of 80\% water and 20\% plastic scintillators, with a mean neutrino energy of 1.5\,GeV~\cite{cite:koga} have also been measured. However, there has been only one publication of CC0$\pi$ anti-neutrino cross-sections on water using P{\O}D~\cite{cite:tn328} with a neutrino energy peak at 0.6 GeV. In this article, we measure CC0$\pi$0p (CC0pi without detected protons) cross-sections on water and hydrocarbon in anti-neutrino beam mode by using a new neutrino detector called the WAGASCI module~\cite{cite:wagasci}, and other T2K detectors, the Proton Module~\cite{cite:pm} and the INGRID module~\cite{cite:ingrid} with a mean neutrino energy of 0.86 GeV at an off-axis angle of 1.5 degrees. As described in Sec.~\ref{sec:detconfig}, the WAGASCI module and the INGRID Water Module are basically the same except for the detector position and electronics. In the future, we will use both detectors to measure neutrino cross-sections at an off-axis angle 1.5 degrees.

%% Paper introduction
Hereafter, we will describe the experimental apparatus, the Monte Carlo simulations, the datasets,  the event selections, the analysis method, the systematic uncertainties, and the results.

% SECTION: EXPERIMENTAL APPARATUS
\section{Experimental Apparatus}\label{sec:experimental_apparatus}
%% Neutrino Beam
\subsection{Neutrino Beam}
The accelerator complex J-PARC in Tokai (Japan) is composed of a linear accelerator (LINAC), a rapid cycling synchrotron (RCS), and the main ring (MR).  The 30\,GeV proton beam is extracted from the MR every 2.48 s. The beam spill consists of eight bunches with 581 ns interval. The protons impinge onto a graphite target fixed in the most upstream electromagnetic horn. Produced charged hadrons are focused by three electromagnetic horns into a 96 m-long decay volume where they decay preferentially producing \nm ($\overline{\nu}_{\mu}$) and $\mu^{+}$ ($\mu^{-}$). By changing the polarity of the horns, the beam mode can be switched between the neutrino mode and anti-neutrino mode. In this article, the data are collected in the anti-neutrino mode with a beam power of about 470 kW.

%% Detector Configuration
\subsection{Detector Configuration}\label{sec:detconfig}
\begin{figure}[!h]
\centering
\includegraphics[width=1.0\linewidth]{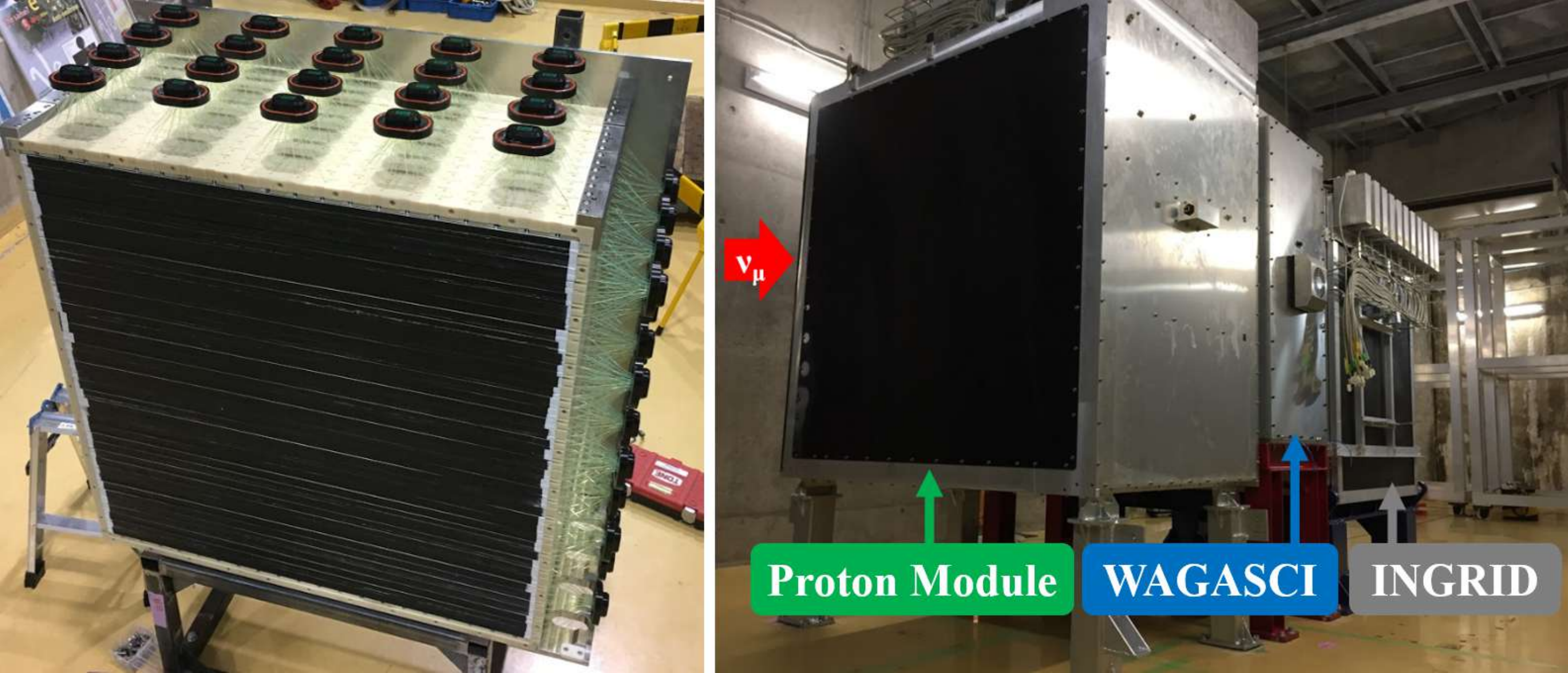}
\caption{The WAGASCI module before the installation into the water tank is shown on the left side. Detectors installed at the J-PARC neutrino-monitor building (right).}
\label{fig:3detectors}
\end{figure}

\begin{figure}[!h]
\centering
\includegraphics[width=0.8\linewidth]{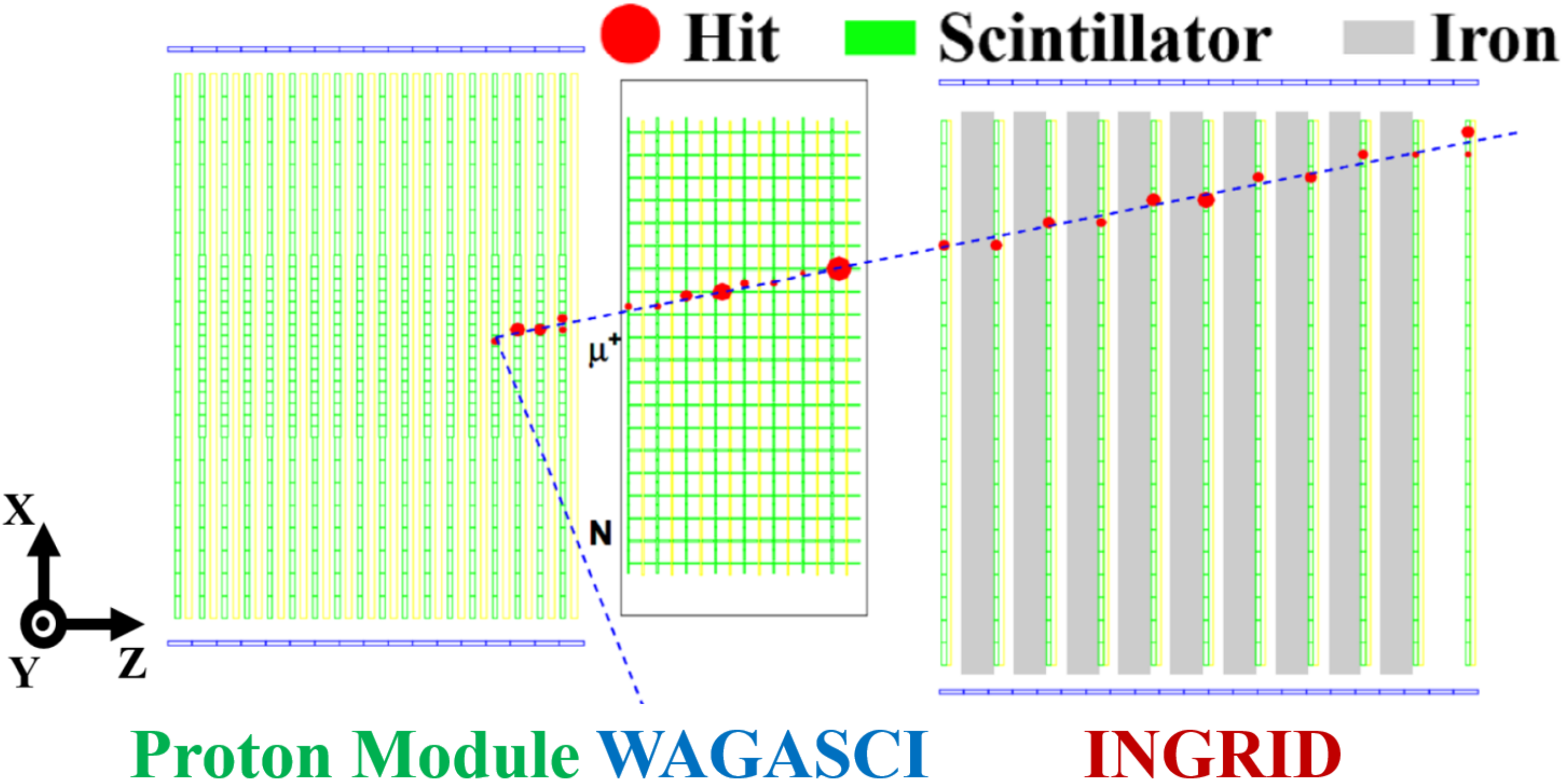}
\caption{Typical neutrino event display for a simulated neutrino event in the Proton Module. The beam axis corresponds to the {\it z}-axis. The muon angle is defined as the scattering angle with respect to {\it z}-axis.}
\label{fig:typical_event_display}
\end{figure}

We use two detectors with different interaction targets, the WAGASCI module (water) and the Proton Module (hydrocarbon). The INGRID module is located at the most downstream position as shown in Fig.~\ref{fig:3detectors}, and is used as a muon detector. These detectors are located at an off-axis angle of 1.5 degrees in the T2K near-detector hall since August 2017. They are exposed to neutrinos with a higher energy distribution than the ND280 detector, since the off-axis angle is smaller than the ND280 angle of 2.5 degrees. A typical event display is shown in Fig.~\ref{fig:typical_event_display}.

\begin{figure}[!h]
\centering
\includegraphics[width=1.0\linewidth]{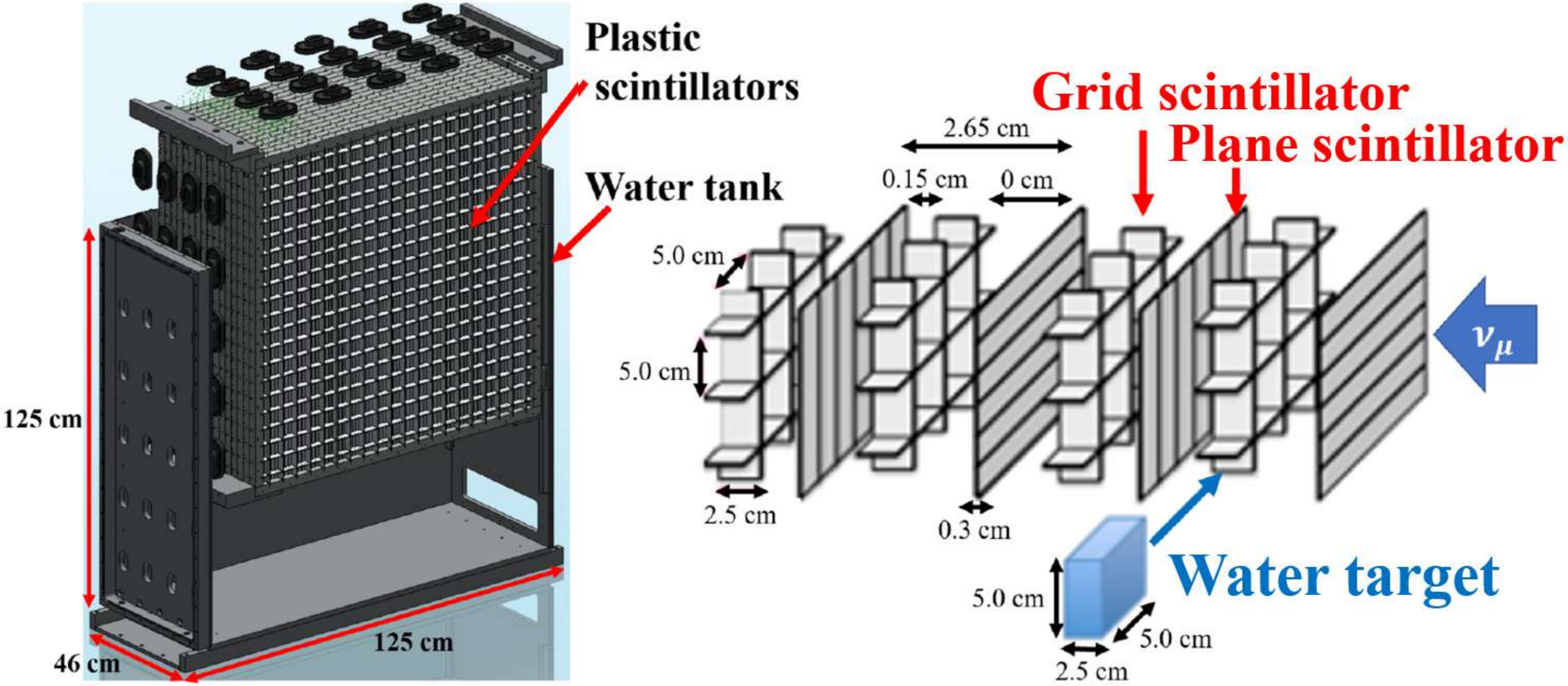}
\caption{Schematic view of the WAGASCI module (left) and its scintillator structure (right).}
\label{fig:grid_structure}
\end{figure}

\begin{figure}[!h]
\centering
\includegraphics[width=0.85\linewidth]{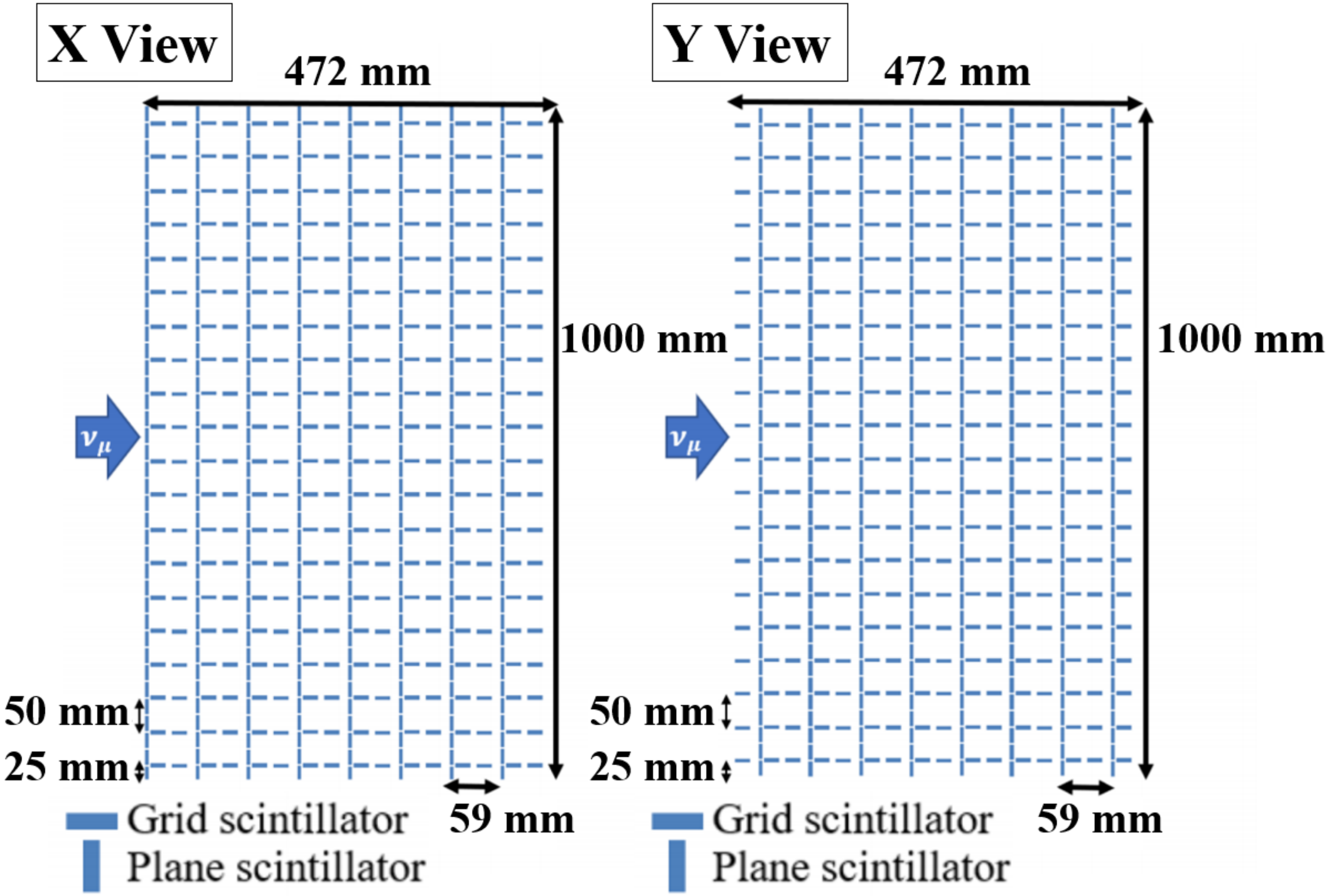}
\caption{View of the scintillators of the WAGASCI module from the {\it x}-direction (left) and {\it y}-direction (right).}
\label{fig:scintillator_view}
\end{figure}

%% WAGASCI
The WAGASCI module is a neutrino detector with 0.6 tons of water and 1280 plastic scintillator bars. The total fraction of water target in the fiducial volume is 80\%, and is higher than the one in other T2K detectors (P{\O}D and FGD2)~\cite{cite:fgd}. The type of scintillator bar ($3\times25\times1020~\rm{cm^{3}}$) and wavelength-shifting (WLS) fiber (Kuraray, Y-11(200)) used in the WAGASCI module is the same as that used in the INGRID Water Module~\cite{cite:koga}. The readout electronics are newly developed with a Silicon PM Integrated Read-Out Chip (SPIROC) which is a 36-channel auto-triggered front-end ASIC. The WAGASCI module consists of 16 scintillator tracking planes in total, and each tracking plane consists of 40 scintillators positioned perpendicularly to the neutrino-beam axis (plane scintillator) and another 40 scintillators positioned in parallel to the beam with a grid structure (grid scintillator), as shown in Fig.~\ref{fig:grid_structure}. Figure~\ref{fig:scintillator_view} shows the schematic view of the scintillators from the {\it x}- and {\it y}-directions, where the definition of the coordinate system is shown in Fig.~\ref{fig:typical_event_display}.

\begin{figure}[!h]
\centering
\includegraphics[width=0.85\linewidth]{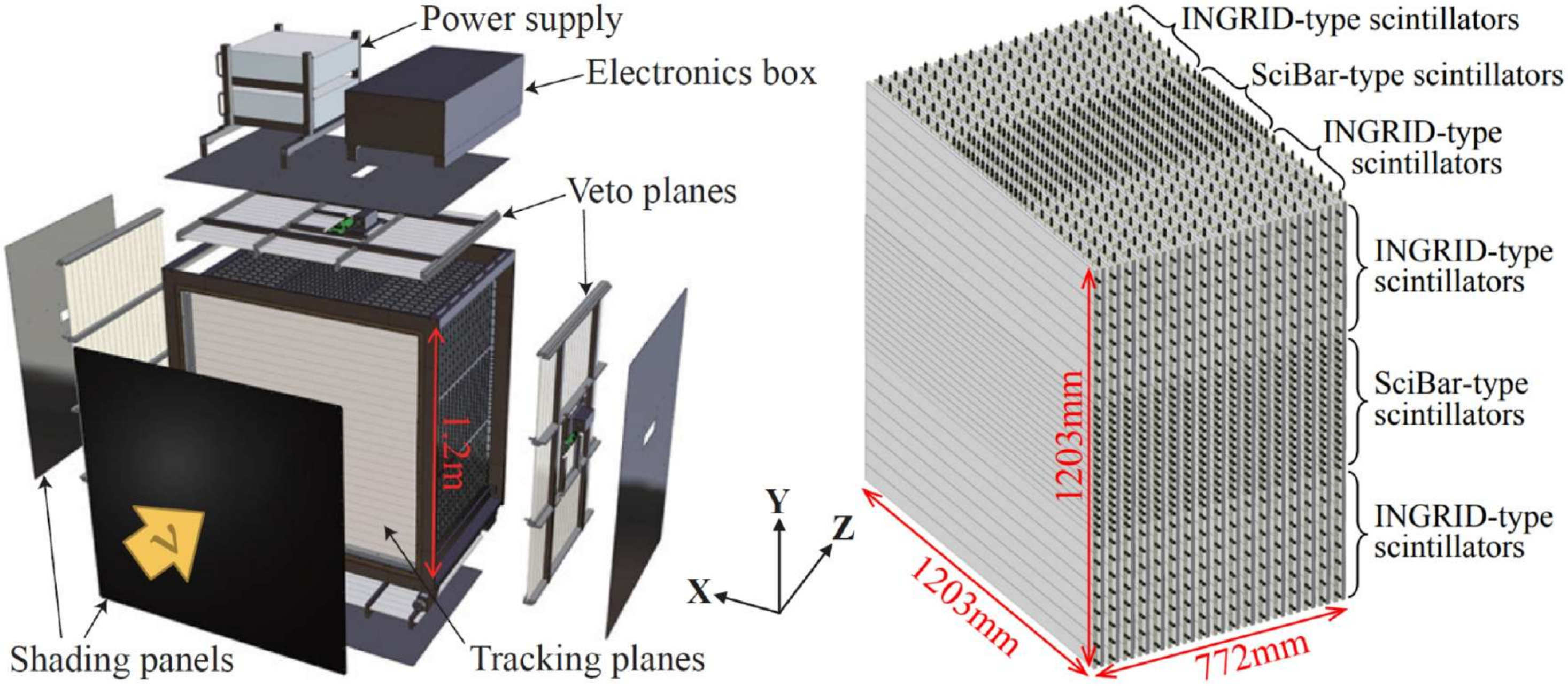}
\caption{Schematic view of the Proton Module (left) and its scintillator structure (right).}
\label{fig:proton_module}
\end{figure}

%% Proton Module
The Proton Module is a fully active tracking detector. It consists of 34 tracking planes, where each tracking plane is an array of two types of 32 scintillator bars, as shown in Fig.~\ref{fig:proton_module}. Two types of scintillators, SciBar type ($13\times25\times1203~\rm{cm^{3}}$) and INGRID type ($10\times50\times1203~\rm{cm^{3}}$), are used, and their chemical composition is the same as that of the WAGASCI type scintillator bar. The six veto planes surrounding the tracking planes are used to track the charged particles coming from outside the Proton Module. The tracking planes also serve as the neutrino-interaction target. The target mass in the fiducial volume is 303 kg in total which corresponds to 98\% of the total target mass. More detailed information about the Proton Module can be found in Ref.~\cite{cite:pm}.

\begin{figure}[!h]
\centering
\includegraphics[width=0.85\linewidth]{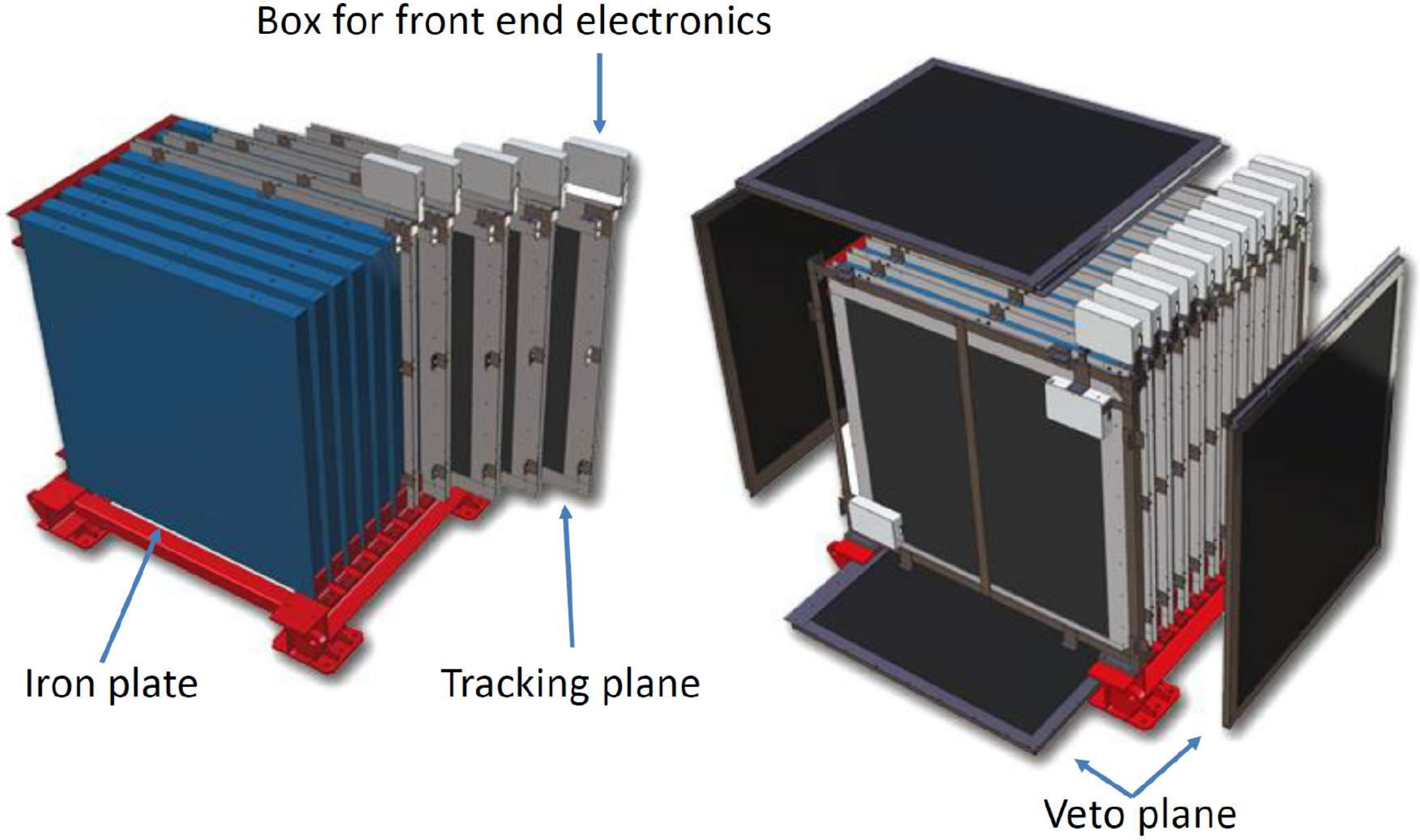}
\caption{Schematic view of the INGRID module.}
\label{fig:ingrid}
\end{figure}

%% INGRID
The INGRID module has a sandwich structure comprising nine iron plates and eleven tracking planes which are surrounded by veto planes, as shown in Fig.~\ref{fig:ingrid}. The tracking planes are formed by two scintillator layers each of which is composed of 24 scintillator bars oriented perpendicularly to one another. The thicknesses of each iron plate and scintillator bar are 6.5\,cm and 1.0\,cm, respectively. More detailed information about the INGRID module can be found in Ref.~\cite{cite:ingrid}.

%% Electronics-like
In all three detectors, the scintillation light emitted from the scintillator bar is collected by a WLS fiber, and it is detected by a Multi-Pixel Photon Counter (MPPC)~\cite{cite:mppc}. To digitize and record the integrated charge and hit timing of 1280 channels, the SPIROC2D electronics~\cite{cite:wagasci_electronics} are used for the WAGASCI module, and the Trip-t electronics~\cite{cite:tript} are used for the Proton Module and the INGRID module. For each beam bunch, the threshold is set to 2.5 p.e. (photon equivalent) to exclude accidental dark noise from MPPCs.

%% FIG: WAGASCI

%% FIG: Proton Module

%% FIG: INGRID

% SECTION: MONTE-CARLO SIMULATION %%%
\section{Monte Carlo Simulation}\label{sec:monte_carlo_simulation}
\begin{figure}[!h]
\begin{minipage}{0.5\hsize}
\centering
\includegraphics[width=1.0\linewidth]{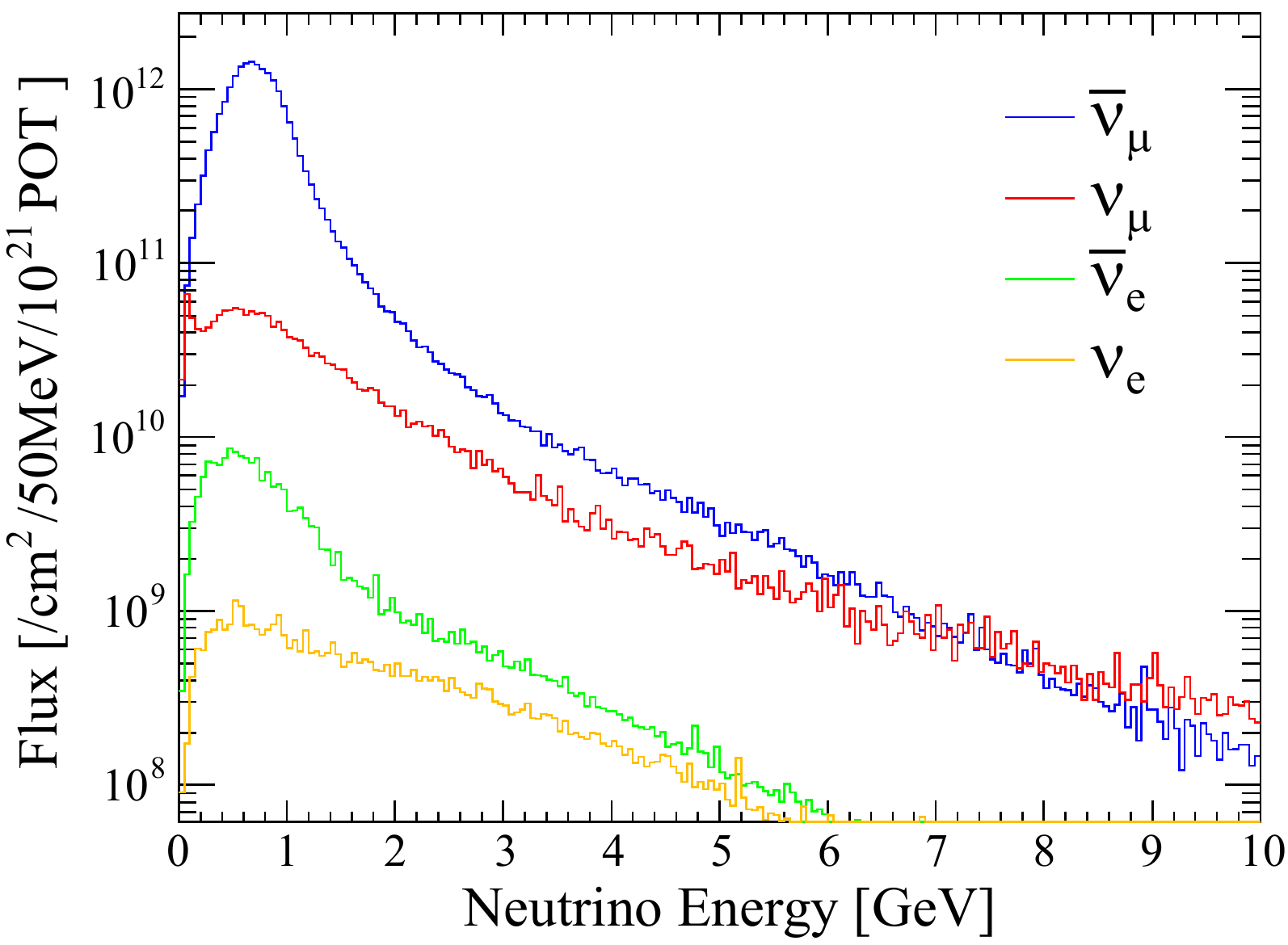}
\end{minipage}
\begin{minipage}{0.5\hsize}
\centering
\includegraphics[width=1.0\linewidth]{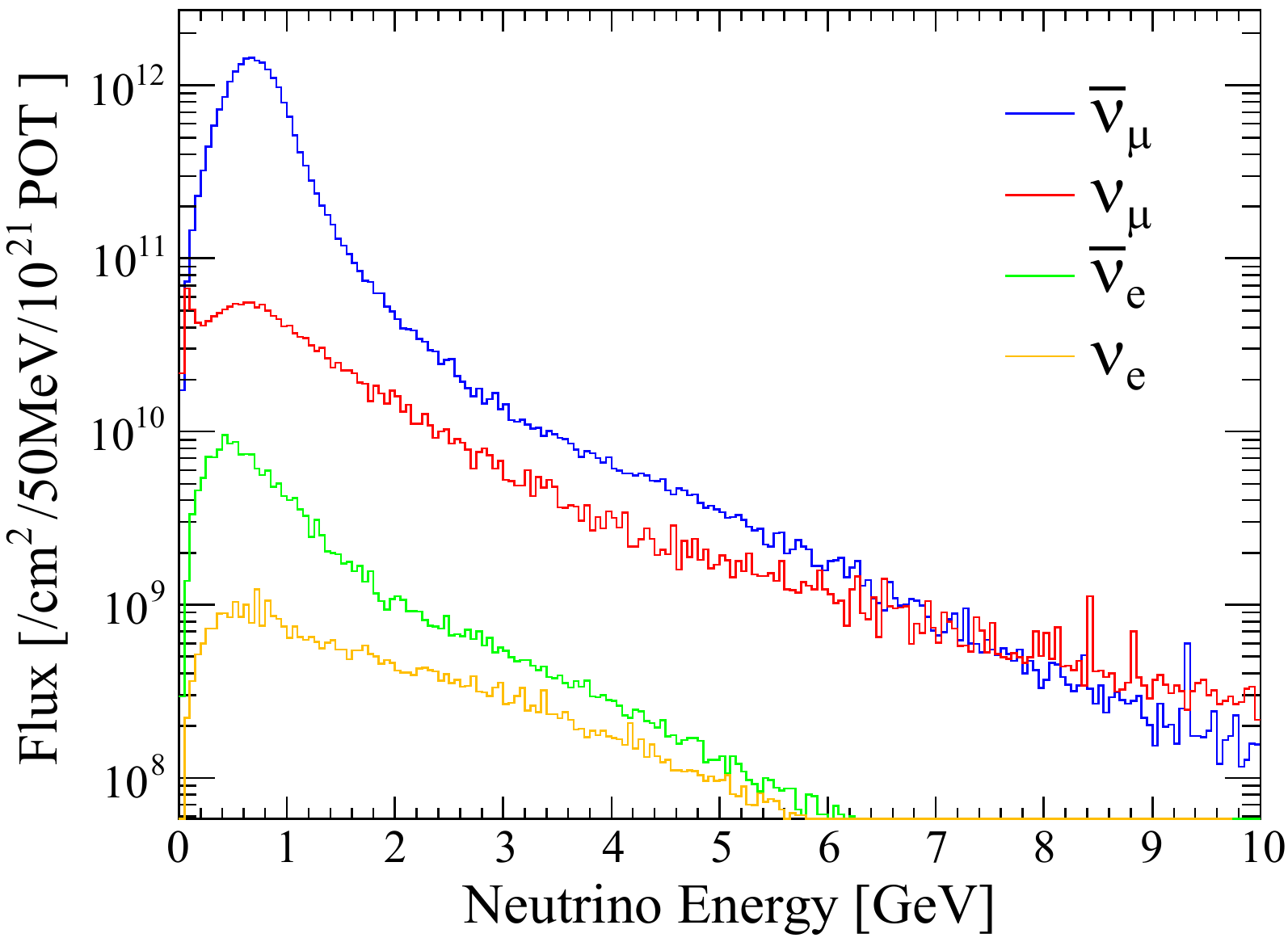}
\end{minipage}
\caption{Predicted (anti-)neutrino fluxes at the WAGASCI module (left) and the Proton Module (right).}
\label{fig:nubeam_flux}
\end{figure}

To estimate backgrounds, neutrino fluxes and signal detection efficiencies, a set of Monte Carlo (MC) simulations is used as follows:

\begin{itemize}
    \item JNUBEAM~\cite{cite:jnubeam} (13a) for neutrino fluxes,
    \item NEUT~\cite{cite:neut} (5.3.3) for neutrino interactions with nuclei,
    \item GEANT4~\cite{cite:geant4}(v9r2p01n00)-based software for the transport and detection of secondary particles.
\end{itemize}
Software settings for the simulation are the same as those used in~\cite{cite:koga}. The neutrino energy spectra at the WAGASCI and Proton Module positions predicted by JNUBEAM, with hadronic processes tuned from the NA61/SHINE measurements~\cite{cite:na61shine}, are shown in Fig.~\ref{fig:nubeam_flux}. The mean neutrino energy is 0.86~GeV, and the peak is at 0.66~GeV with 1$\sigma$ spread of $^{+0.40}_{-0.25}$~GeV. The flux-integrated CC cross-sections per nucleon predicted by NEUT are summarized in Table~\ref{tab:neut_expected_xsec}.
To compare predicted neutrino cross-sections in Sec.~\ref{sec:result}, an alternative event generator, GENIE~\cite{cite:genie} (2.12.8), is also used. In both generators, a Relativistic Fermi-Gas (RFG) model~\cite{cite:rfg} is used, but the Bodek-Ritchie modifications~\cite{cite:bodek_ritchie1, cite:bodek_ritchie2} are implemented in GENIE. In NEUT, random-phase approximation (RPA)~\cite{cite:rpa}, and multi-nucleon (2p2h) interactions~\cite{cite:2p2h} are considered. In addition, they use the Rein-Sehgal model~\cite{cite:rs1, cite:rs2} for the single-meson production, the Berger-Sehgal model~\cite{cite:bs} for the coherent-pion production, and $\rm Gl\ddot{u}ck$-Reya-Vogt-1998 (GRV98)~\cite{cite:grv98} parton distributions with Bodek-Yang modifications~\cite{cite:by1, cite:by2} for the deep-inelastic scattering. NEUT is also used for the T2K neutrino oscillation analysis, and more details can be found in Ref.~\cite{cite:t2k_neutrino_oscillation}. %The number of generated MC events corresponds to those with $1.0 \times 10^{24}$ POT.}

%% Flux

\begin{table}[!h]
\centering
\caption{Flux-integrated \nmb CC0$\pi$0p cross-sections per nucleon on $\rm H_{2}O$ and CH predicted by NEUT.}
\label{tab:neut_expected_xsec}
\begin{tabular}{ccc}
\hline\hline
Cross section                                       & NEUT expectation with RPA             & NEUT expectation without RPA \\
\hline
$\sigma_{\rm H_{2}O}$                            & $1.013 \times 10^{-39}~{\rm cm}^{2}$ & $1.189 \times 10^{-39}~{\rm cm}^{2}$ \\
$\sigma_{\rm CH}$                                & $1.051 \times 10^{-39}~{\rm cm}^{2}$ & $1.278 \times 10^{-39}~{\rm cm}^{2}$ \\
$\sigma_{\rm H_{2}O}$/$\sigma_{\rm CH}$ & 0.964                                           & 0.930 \\
\hline\hline
\end{tabular}
\end{table}

% SECTION: DATASETS & EVENT SELECTIONS %%%
\section{Datasets and Event Selections}\label{sec:event_selection}
In this article, data collected from October 2017 to May 2018 are used. The datasets include statistics of $7.91\times10^{20}$ POT in the anti-neutrino mode. The signal events in the WAGASCI module and the Proton Module are selected from these data. In this analysis, the signal is defined as the charged-current interaction with no detected pions nor protons. This signal is characterized by a muon-like track produced inside the detector. The cross-section is calculated for signal events both from \nmb interactions (\nmb cross-section) and \nmbm interactions (\nmbm cross-section), as described in Sec. \ref{sec:numu+numubar}.

%% Event selections
The selections applied to the two detectors are similar to those in a previous analysis~\cite{cite:koga}, where cross-sections on water and hydrocarbon targets were measured. The selection criteria in this analysis are briefly described below.

% For WAGASCI
\subsection{Selections for the WAGASCI module}

% Event reconstruction
A scintillator channel having an ADC charge greater than 2.5 p.e. is defined as a ``hit''. Based on a cellular automaton algorithm~\cite{cite:cellular_automaton}, these hits are fitted by a line (track reconstruction). The two-dimensional tracks are reconstructed in each detector from more than two hits in a beam bunch, and then at least one track in the WAGASCI module and the Proton Module is required to be matched with a reconstructed track in the INGRID module to select a muon-like track.
Three-dimensional tracks are searched for among pairs of two-dimensional XZ tracks and YZ tracks.
After the reconstruction of the three-dimensional tracks, the upstream point of the longest track is defined as a neutrino interaction vertex.

% Other cuts
Subsequently, in order to reduce non-beam backgrounds such as cosmic rays, the event timing for a vertex is required to be within 100 ns from the expected beam-bunch timing (beam-timing cut). In addition, to reduce the beam-induced backgrounds mainly from neutrino interactions in the walls of the detector hall, two cuts are applied. First, if the most upstream point of a reconstructed track is in the first or second plane of the parallel scintillators, then that event is excluded. Second, if a vertex is in the outer region of the fiducial volume (FV), then that event is excluded. The FV is defined as the central area of the WAGASCI module with dimensions of 70 cm (in {\it x}-coordinate) $\times$ 70 cm (in {\it y}-coordinate) $\times$ 21 cm (in {\it z}-coordinate).

\begin{figure}[!h]
\centering
\includegraphics[width=0.7\linewidth]{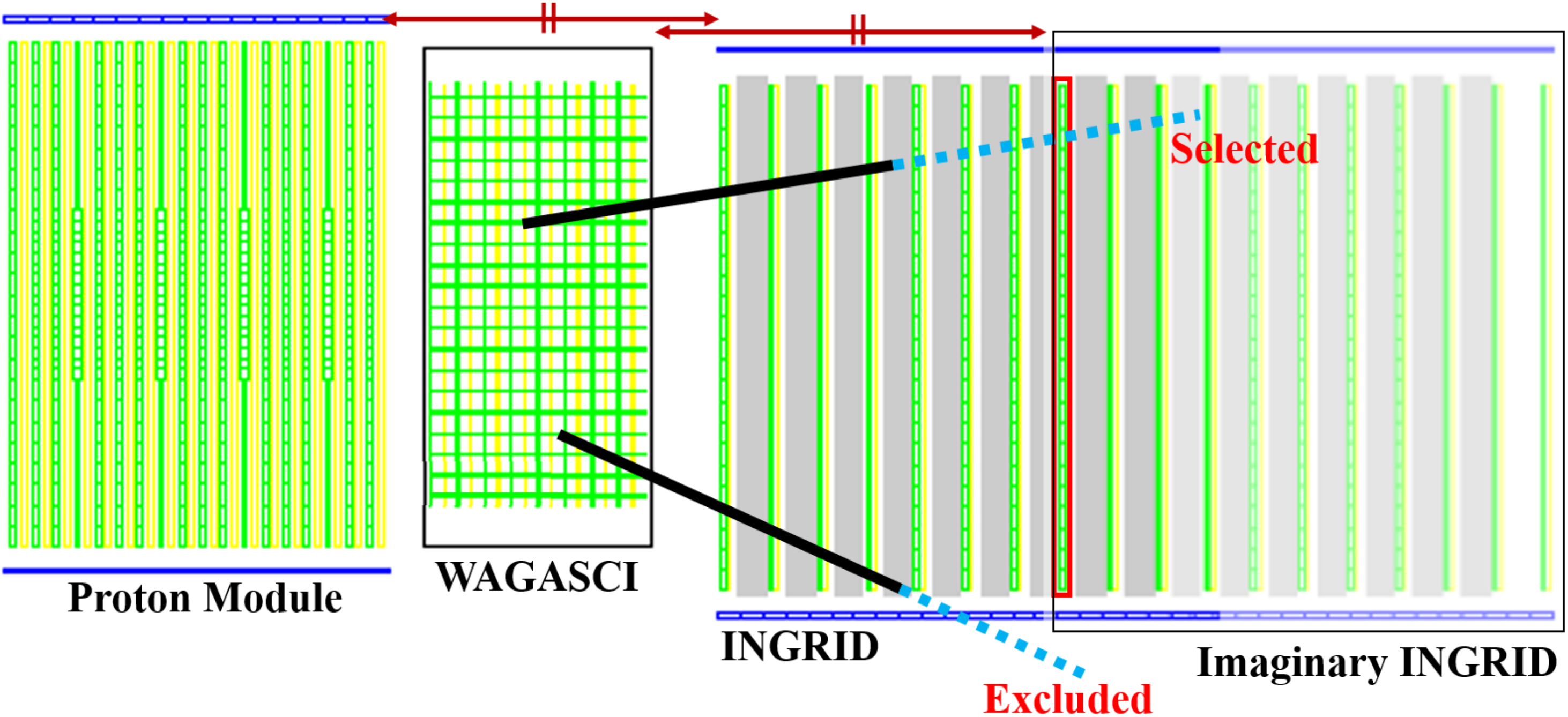}
\caption{Schematic view of selected and excluded events by the angular acceptance cut for the WAGASCI module. If the extrapolated track from the WAGASCI module reaches the imaginary INGRID module, the event is selected.}
\label{fig:imaginary_ingrid}
\end{figure}

% Acceptance cut
Since the WAGASCI module lies closer to the INGRID module than the Proton Module, the angular acceptance by the INGRID module is larger.
In order to obtain a similar angular acceptance to that of the Proton Module, an extrapolation of the reconstructed track from the WAGASCI module is required to reach an imaginary INGRID module. The imaginary INGRID module is set as shown in Fig.~\ref{fig:imaginary_ingrid} so that the distance between the downstream edge of the Proton Module and the upstream edge of the INGRID module (1034.5 cm) is almost the same as that between the downstream edge of the WAGASCI module and the upstream edge of the imaginary INGRID module (1035.5 cm).

%% FIG

% One-track extraction
For signal interactions, a single muon-like track is expected in the final state. To reduce the multi-track backgrounds from other neutrino interactions, events having more than one track are excluded.

\begin{figure}[!h]
  \centering
  \includegraphics[width=1.0\linewidth]{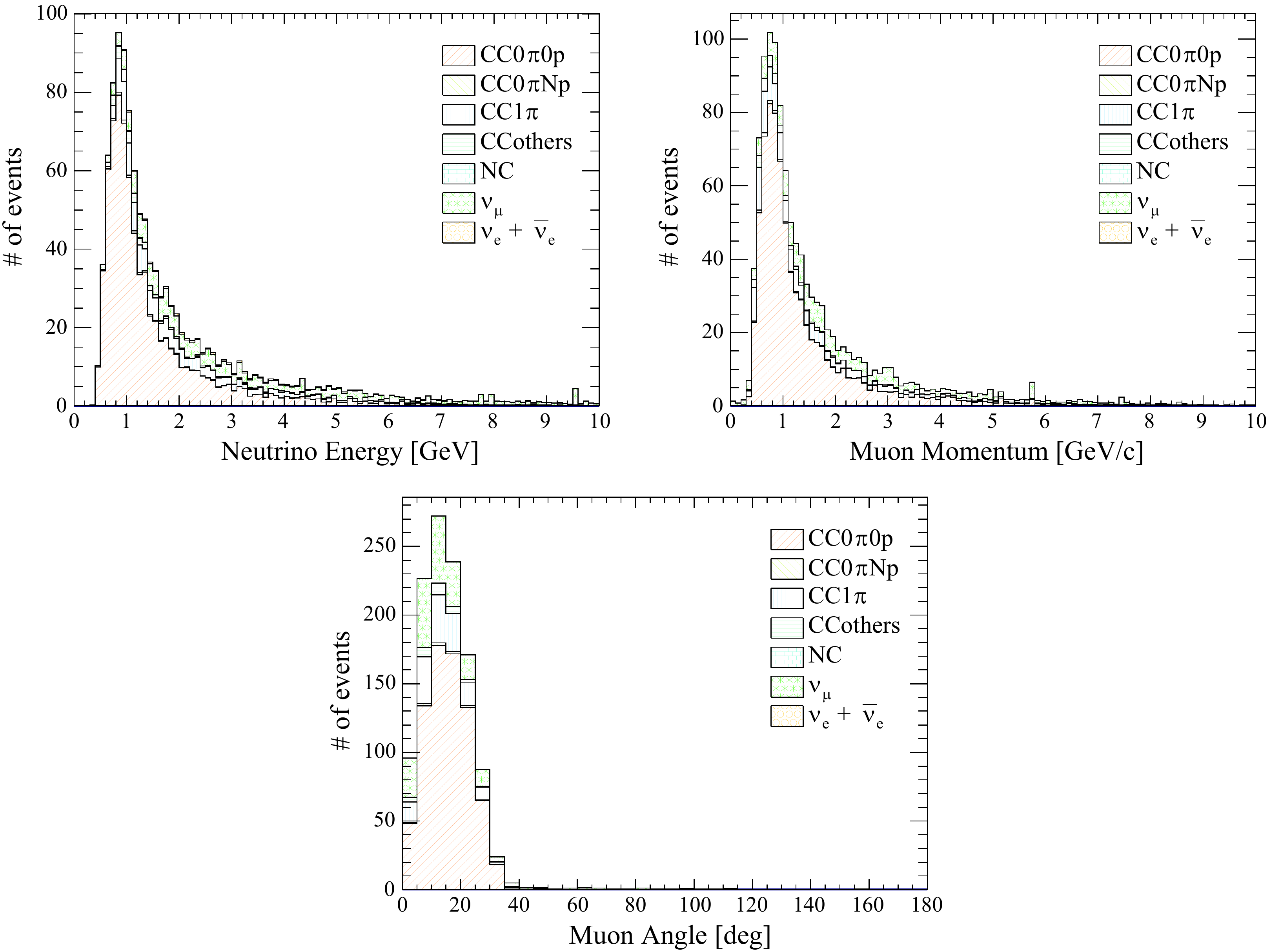}
  \caption{True distribution of the selected events in the WAGASCI module as a function of neutrino energy (top left), muon momentum (top right), and muon angle (bottom).}
  \label{fig:selected_events_wm}
\end{figure}

\begin{figure}[!h]
\begin{minipage}{0.5\hsize}
\centering
\includegraphics[width=1.0\linewidth]{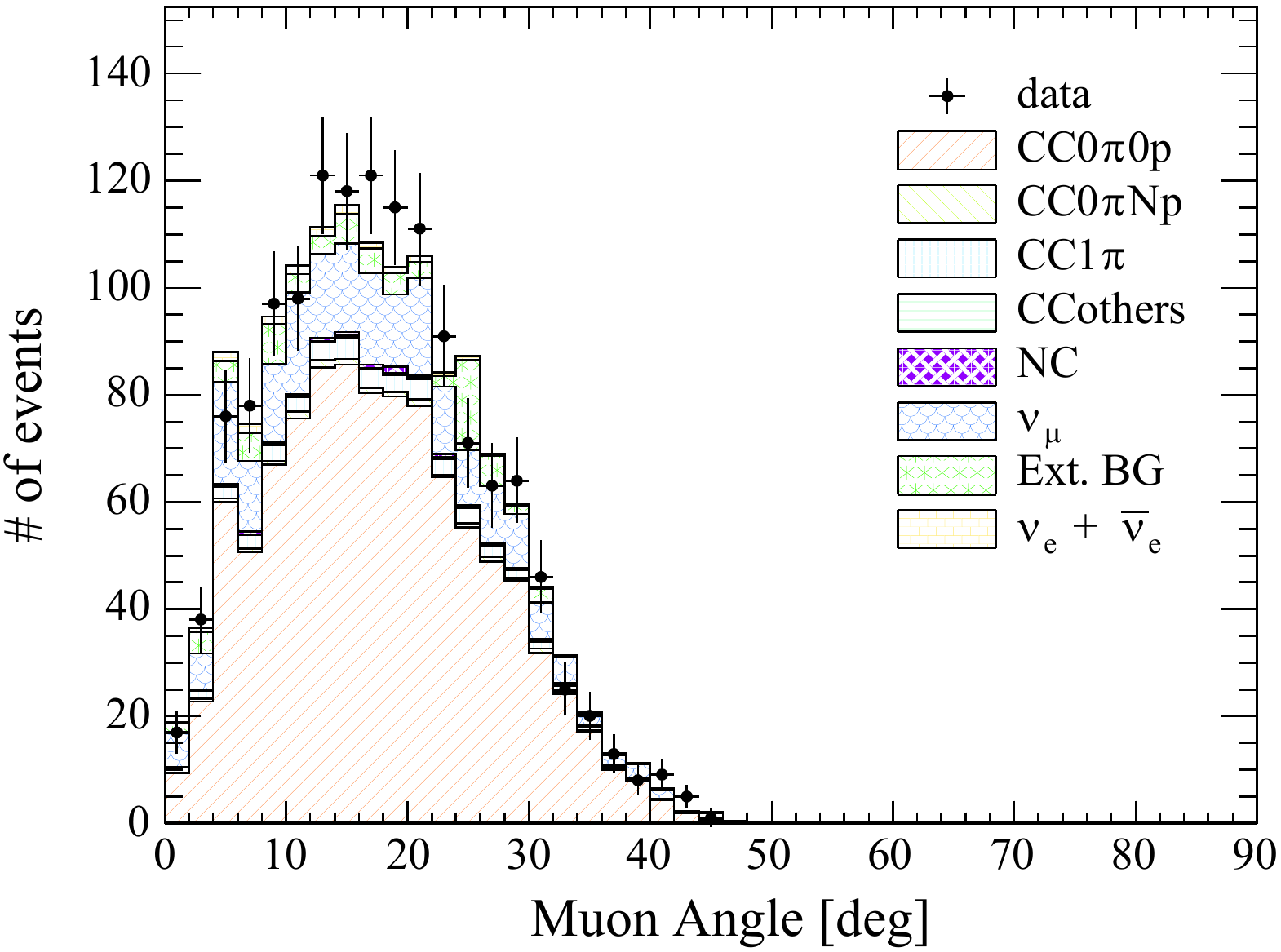}
\end{minipage}
\begin{minipage}{0.5\hsize}
\centering
\includegraphics[width=1.0\linewidth]{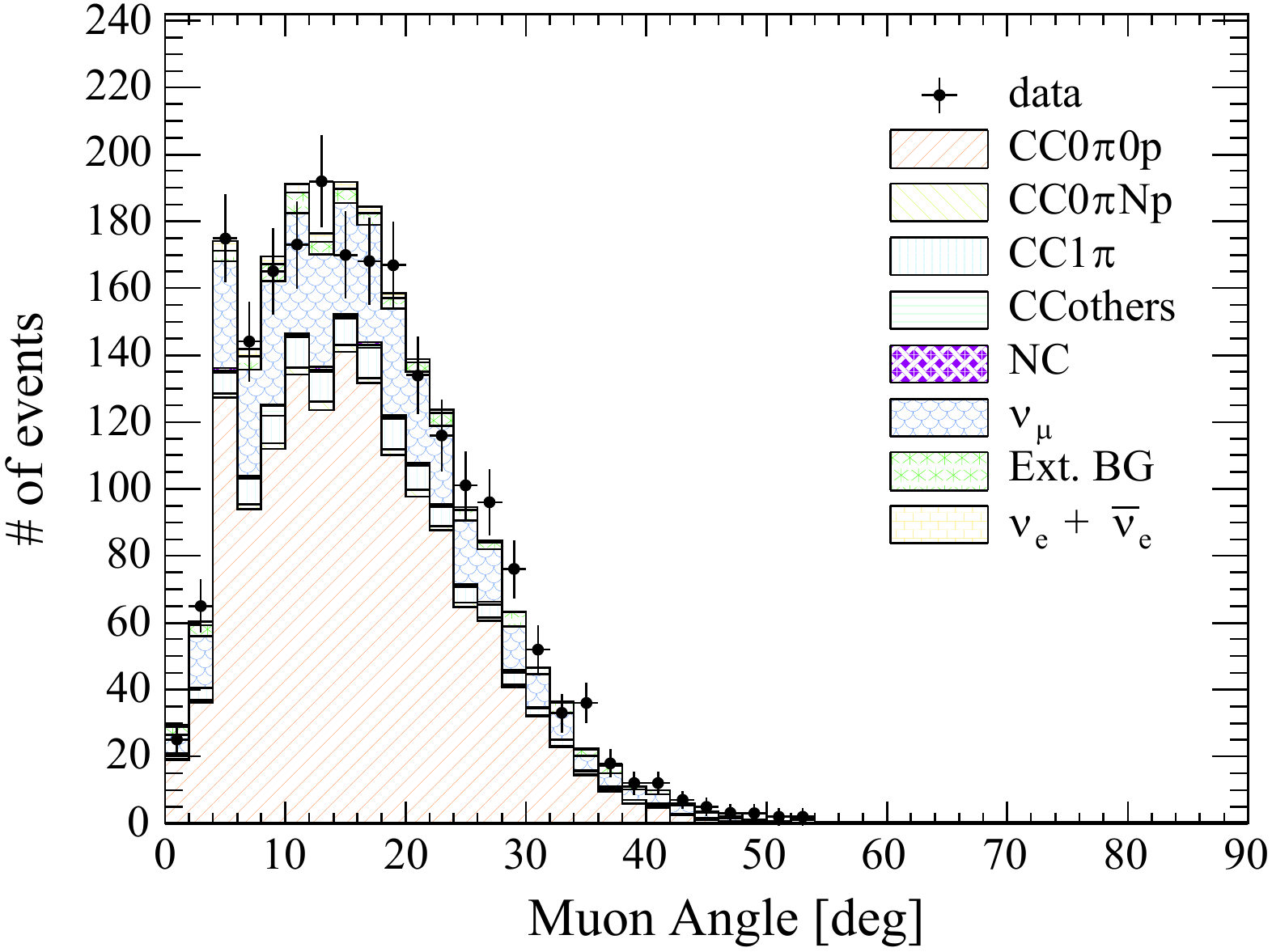}
\end{minipage}
\caption{Angular distributions of the longest reconstructed track matched with the INGRID track after the one-track extraction for the WAGASCI module (left) and the Proton Module (right).}
\label{fig:reconangle}
\end{figure}

% Summary for WAGASCI
The number of selected events and the background fraction in the WAGASCI module are summarized in Table~\ref{tab:event_summary_WM}. The last cut of the reconstructed track angle is due to the final selection acceptance, and it is described in Sec.~\ref{sec:selection_efficiency}. The neutrino energy, muon momentum, and angular distributions of the selected events predicted by the MC simulation are shown in Fig.~\ref{fig:selected_events_wm}. The left panel of Fig.~\ref{fig:reconangle} shows the angular distribution of the reconstructed single muon-like track for events passing the one-track extraction in the WAGASCI module.

\begin{table}[!h]
\centering
\caption{Summary of event selections for the WAGASCI module. The numbers written in brackets represent the fraction of the total number of events passing each selection. The number of events predicted by MC is normalized to the actual recorded POT ($7.9 \times 10^{20}$).}
\label{tab:event_summary_WM}
\scalebox{0.85}{
\begin{tabular}{l|ccccc|cc}
\hline\hline
\multirow{2}{*}{Selection}    & \multicolumn{5}{c|}{MC} & \multirow{2}{*}{Data} & \multirow{2}{*}{Data/MC} \\
                                     & \nmb & \nm & \ne~+~\neb & External B.G. & Total & & \\
\hline  
\multirow{2}{*}{Event reconstruction}          & 5559.1   & 2597.9   & 149.9   & 10582.9  & 18889.7   & 20728 & 1.10 \\
                                     & (29.4\%) & (13.8\%) & (0.8\%) & (56.0\%) & (100.0\%) &  &  \\
\hline
\multirow{2}{*}{Beam timing}                    & 5485.5   & 2462.8   & 142.3   & 10439.1  & 18529.7   & 20095 & 1.08 \\
                                    & (29.6\%) & (13.3\%) & (0.8\%) & (56.3\%) & (100.0\%) &  &  \\
\hline
\multirow{2}{*}{Upstream veto}                & 3925.3   & 1755.0   & 83.0    & 6081.8   & 11845.1   & 12236 & 1.03 \\
                                    & (33.1\%) & (14.8\%) & (0.7\%) & (51.3\%) & (100.0\%) &  &  \\
\hline
\multirow{2}{*}{Fiducial volume}               & 1936.9   & 812.8    & 38.7    & 112.3    & 2900.7    & 2797 & 0.96 \\
                                    & (66.8\%) & (28.0\%) & (1.3\%) & (3.9\%)  & (100.0\%) &  &  \\
\hline
\multirow{2}{*}{Additional acceptance}       & 1279.9   & 497.4    & 28.3    & 81.5     & 1887.1    & 1783 & 0.94 \\
                                    & (67.8\%) & (26.4\%) & (1.5\%) & (4.3\%)  & (100.0\%) &  &  \\
\hline
\multirow{2}{*}{One-track extraction}            & 1075.7   & 224.5    & 17.3    & 76.5     & 1394.0    & 1406 & 1.01 \\
                                    & (77.2\%) & (16.1\%) & (1.2\%) & (5.5\%)  & (100.0\%) &  &  \\
\hline
\multirow{2}{*}{Reconstructed track angle} & 969.5    & 203.5    & 16.5    & 72.3     & 1261.9    & 1279 & 1.01 \\
                                    & (76.8\%) & (16.1\%) & (1.3\%) & (5.7\%)  & (100.0\%) &  &  \\
\hline
\hline
\end{tabular}
}
\end{table}

% For Proton Module
\subsection{Selections for the Proton Module}
Selection criteria for the Proton Module basically use the same method as those for the WAGASCI module, except for the two-dimensional track matching. Since the WAGASCI module is located between the Proton Module and the INGRID module, two-dimensional tracks in the Proton Module are required to be matched to both the WAGASCI module and the INGRID module.

\begin{figure}[!h]
  \centering
  \includegraphics[width=1.0\linewidth]{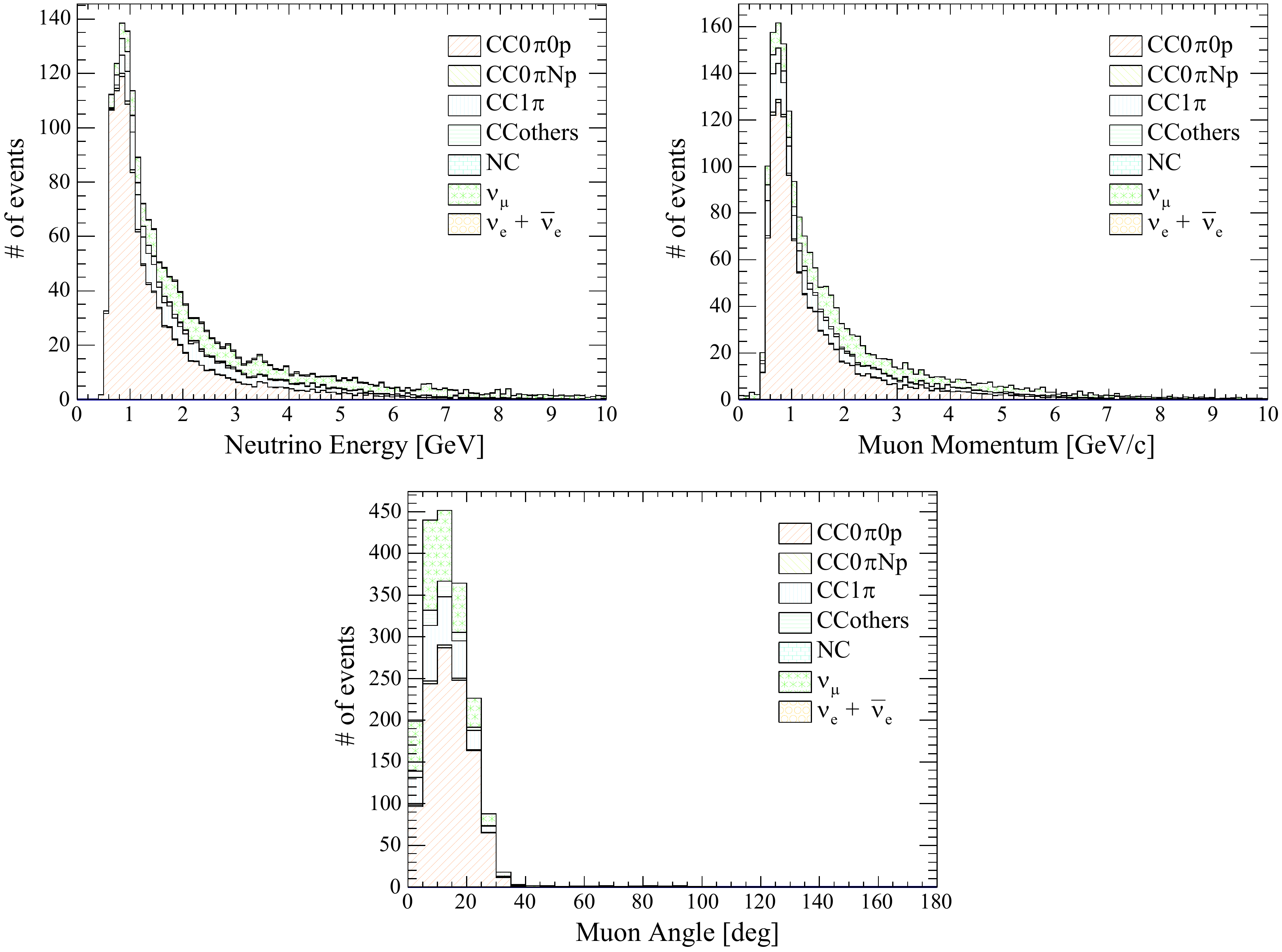}
  \caption{True distribution of the selected events in the Proton Module as a function of neutrino energy (top left), muon momentum (top right), and muon angle (bottom).}
  \label{fig:selected_events_pm}
\end{figure}

% Summary for Proton Module
The number of selected events and the background fraction in the Proton Module are summarized in Table~\ref{tab:event_summary_PM}. The neutrino energy, muon momentum, and angular distributions of the selected events predicted by the MC simulation are shown in Fig.~\ref{fig:selected_events_pm}. The right panel of Fig.~\ref{fig:reconangle} shows the angular distribution of the reconstructed single muon-like track for events passing the one-track extraction in the Proton Module.

\begin{table}[!h]
\centering
\caption{Summary of event selections for the Proton Module. The numbers written in brackets represent the fraction in the total number of events passing each selection. The number of events predicted by MC is normalized to the actual recorded POT ($7.9 \times 10^{20}$).}
\label{tab:event_summary_PM}
\scalebox{0.85}{
\begin{tabular}{l|ccccc|cc}
\hline\hline
\multirow{2}{*}{Selection}    & \multicolumn{5}{c|}{MC} & \multirow{2}{*}{Data} & \multirow{2}{*}{Data/MC} \\
                                     & \nmb & \nm & \ne~+~\neb & External B.G. & Total & & \\
\hline  
\multirow{2}{*}{Event reconstruction}          & 4813.4   & 2219.1   & 104.1   & 195761.9  & 202898.5   & 191554 & 0.94 \\
                                                          & (2.4\%) & (1.1\%) & (0.1\%) & (96.5\%) & (100.0\%) &  &  \\
\hline
\multirow{2}{*}{Beam timing}                    & 4807.8   & 2201.8   & 103.3   & 195691.1  & 202804.0   & 191118 & 0.94 \\
                                                         & (29.6\%) & (13.3\%) & (0.8\%) & (56.3\%) & (100.0\%) &  &  \\
\hline
\multirow{2}{*}{Upstream veto}                & 4223.2   & 1883.3   & 88.6    & 31118.6   & 37313.7   & 40593 & 1.09 \\
                                                         & (11.3\%) & (5.0\%) & (0.2\%) & (83.4\%) & (100.0\%) &  &  \\
\hline
\multirow{2}{*}{Fiducial volume}               & 1865.8   & 792.0    & 39.0    & 71.3    & 2768.2    & 2623 & 0.95 \\
                                                        & (67.4\%) & (28.6\%) & (1.4\%) & (2.6\%)  & (100.0\%) &  &  \\
\hline
\multirow{2}{*}{Additional acceptance}       & 1865.8   & 792.0    & 39.0    & 71.3     & 2768.2    & 2623 & 0.95 \\
                                                         & (67.4\%) & (28.6\%) & (1.4\%) & (2.6\%)  & (100.0\%) &  &  \\
\hline
\multirow{2}{*}{One-track extraction}        & 1620.6   & 429.0    & 25.0    & 68.5     & 2143.1    & 2152 & 1.00 \\
                                                         & (75.6\%) & (20.0\%) & (1.2\%) & (3.2\%)  & (100.0\%) &  &  \\
\hline
\multirow{2}{*}{Reconstructed track angle} & 1514.5    & 390.1    & 23.7    & 54.8     & 1983.1    & 1967 & 0.99 \\
                                                         & (76.4\%) & (19.7\%) & (1.2\%) & (2.8\%)  & (100.0\%) &  &  \\
\hline
\hline
\end{tabular}
}
\end{table}

% Selection efficiency
\subsection{Selection Efficiencies}\label{sec:selection_efficiency}
\begin{figure}[!h]
\begin{minipage}{0.5\hsize}
\centering
\includegraphics[width=0.9\linewidth]{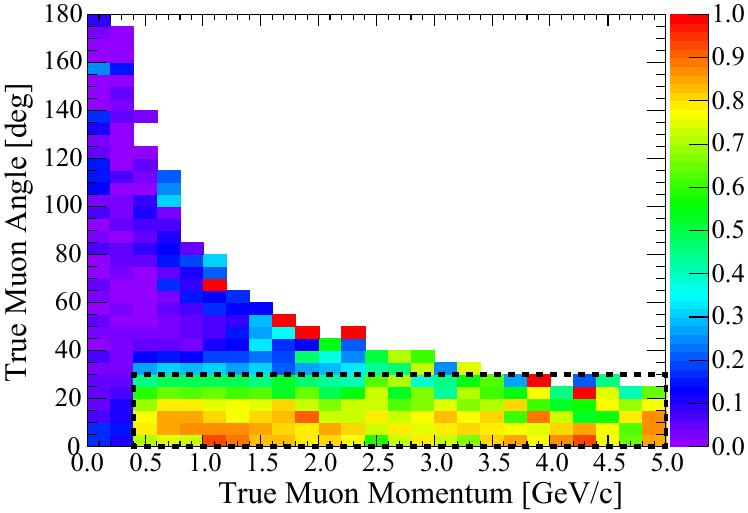}
\end{minipage}
\begin{minipage}{0.5\hsize}
\centering
\includegraphics[width=0.9\linewidth]{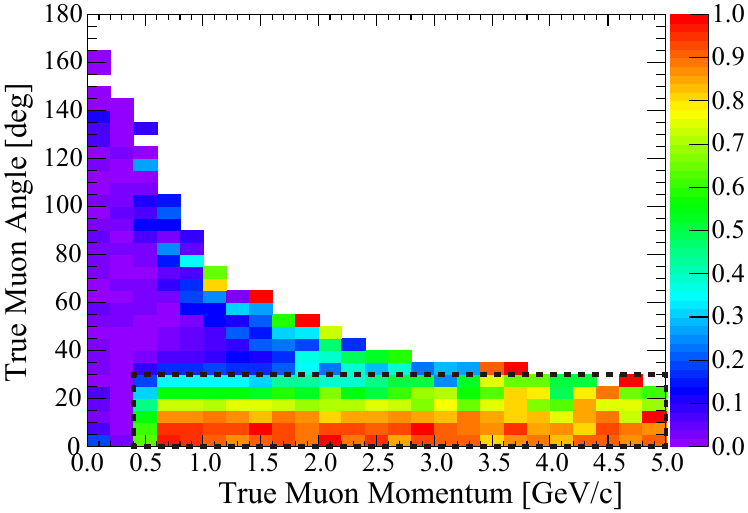}
\end{minipage}
\caption{Detection efficiencies of the WAGASCI module (left) and the Proton Module (right) for \nmb CC events with respect to true muon angle and momentum. The $z$-axis of the plot indicates the detection efficiency. The restricted phase space corresponds to the region inside the dotted square.}
\label{fig:selection_eff}
\end{figure}

Figure~\ref{fig:selection_eff} shows selection efficiencies of CC events for the WAGASCI module and the Proton Module as a function of true muon scattering angle and momentum. The phase spaces of induced muons are restricted to the high-detection efficiency-region, $\theta_{\mu}<30^{\circ}$ and $p_{\mu}>400~{\rm MeV}/c$, in the laboratory frame. According to this restriction, the charged-current events are classified into six bins based on the muon angles, as summarized in Table~\ref{tab:bin_definition}. Although the signal is CC0$\pi$0p with a muon angle smaller than 30 degrees, the selected events for cross-section calculations also include two bins for multi-track samples  (labelled as CCother) and higher angle samples (labelled as single track 30$^{\circ}$-180$^{\circ}$ CC0$\pi$0p). In addition, detectable phase spaces of pions and protons are defined as ``$\theta_{\pi}<70^{\circ}$ and $p_{\pi}>200~{\rm MeV}/c$'' and ``$\theta_{{\rm p}}<70^{\circ}$ and $p_{{\rm p}}>600~{\rm MeV}/c$'', respectively, in the laboratory frame, and the signal phase space is defined allowing no pions nor protons in these regions. Detection efficiencies for each bin are summarized in Table~\ref{tab:selection_efficiency}.

\begin{table}[!h]
\centering
\caption{Bin definition based on true phase space and reconstructed tracks.}
\label{tab:bin_definition}
\begin{tabular}{c|ccc}
\hline\hline
Bin number & Muon angle range & \multicolumn{1}{c}{True phase space} & \multicolumn{1}{c}{Reconstructed track} \\
\hline
0              & 0$^{\circ}$-180$^{\circ}$ & CCother & Multi-track \\
1              & 0$^{\circ}$-5$^{\circ}$ & $\rm CC0\pi0p$ & Single-track \\
2              & 5$^{\circ}$-10$^{\circ}$ & $\rm CC0\pi0p$ & Single-track \\
3              & 10$^{\circ}$-15$^{\circ}$ & $\rm CC0\pi0p$ & Single-track \\
4              & 15$^{\circ}$-20$^{\circ}$ & $\rm CC0\pi0p$ & Single-track \\
5              & 20$^{\circ}$-25$^{\circ}$ & $\rm CC0\pi0p$ & Single-track \\
6              & 25$^{\circ}$-30$^{\circ}$ & $\rm CC0\pi0p$ & Single-track \\
7              & 30$^{\circ}$-180$^{\circ}$ & $\rm CC0\pi0p$ & Single-track \\
\hline\hline
\end{tabular}
\end{table}

\begin{table}[!h]
\centering
\caption{Calculated detection efficiencies of \nmb and \nmbm CC events for each of the phase-space bins.}
\label{tab:selection_efficiency}
\begin{tabular}{l|cc|cc}
\hline
\hline
                                                & \multicolumn{2}{c|}{\nmb}                                  & \multicolumn{2}{c}{\nmbm} \\ \cline{2-5}
True phase space                         & WAGASCI                  & Proton Module             & WAGASCI                    & Proton Module \\
\hline
$\rm CCother$                       & $0.194$       & $0.237$          & $0.233$          & $0.289$            \\
$\rm CC0\pi0p:0^\circ$-$5^\circ$    & $0.683$      & $0.897$          & $0.682$          & $0.897$            \\
$\rm CC0\pi0p:5^\circ$-$10^\circ$   & $0.738$     & $0.896$          & $0.729$          & $0.892$            \\
$\rm CC0\pi0p:10^\circ$-$15^\circ$  & $0.737$    & $0.830$          & $0.724$          & $0.825$            \\
$\rm CC0\pi0p:15^\circ$-$20^\circ$  & $0.679$    & $0.694$          & $0.674$          & $0.693$            \\
$\rm CC0\pi0p:20^\circ$-$25^\circ$  & $0.552$    & $0.502$          & $0.543$          & $0.507$            \\
$\rm CC0\pi0p:25^\circ$-$30^\circ$  & $0.391$    & $0.305$          & $0.387$          & $0.302$            \\
$\rm CC0\pi0p:30^\circ$-$180^\circ$ & $0.081$   & $0.048$          & $0.081$          & $0.048$            \\
\hline
Total                                             & $0.372$    & $0.397$          & $0.355$          & $0.395$  \\
\hline
\hline
\end{tabular}
\end{table}

% SECTION: CROSS-SECTION ANALYSIS %%%
\section{Cross-Section Extraction}\label{sec:xsec_analsis}
%%%%%%%%%%%%%%%%%%%%%%%%%%%%
%%% Cross-section extraction numubar only %%%
%%%%%%%%%%%%%%%%%%%%%%%%%%%%

%%%%%%%%%%%%%%%%%%%%
%%%%%%%%%%%%%%%%%%%%
In this paper, the following notations are used:

\begin{itemize}
	\item $\rb{j}$ represents the {\it j}-th reconstructed single-track angle bin.
	\item $\tb{i}$ represents the {\it i}-th true angle of muons, pions, and protons.
	\item A smearing matrix, $\mathbb{P}(\rb{j}|\tb{i})$, represents a probability that an event from $\tb{i}$ is reconstructed in $\rb{j}$.
	\item An unfolding matrix, $U_{ij}=\mathbb{P}(\tb{i}|\rb{j})$, represents a probability that an event in $\rb{j}$ derives from an event in $\tb{i}$.
\end{itemize}

\noindent The analysis method is almost the same as that used in Ref.~\cite{cite:koga}, and detailed information can be found in that reference.

%%%%%%%%%%%%%%%%%%%%
%%%%%%%%%%%%%%%%%%%%
\subsection{Calculation formula}\label{sec:xsec_extraction_formula}

%%% Formula
The CC0$\pi$0p flux-integrated differential cross-sections are calculated as follows:
\begin{eqnarray}
\label{ math:eq_xsec_h2o }
\sigma_{i~\rm{H_{2}O}} &=& \sum_{j}\frac{ U_{ij~\rm{WM}} (N^{\rm{sel}}_{j~\rm{WM}}-N^{\rm{BG}}_{j~\rm{WM}}) }{\Phi^{\rm{H_{2}O}}_{\rm{WM}}T^{\rm{H_{2}O}}_{\rm{WM}} \varepsilon^{\rm{H_{2}O}}_{i~\rm{WM}}} \\
\label{ math:eq_xsec_ch }
\sigma_{i~\rm{CH}}     &=& \sum_{j}\frac{ U_{ij~\rm{PM}} (N^{\rm{sel}}_{j~\rm{PM}}-N^{\rm{BG}}_{j~\rm{PM}}) }{\Phi^{\rm{CH}}_{\rm{PM}}T^{\rm{CH}}_{\rm{PM}} \varepsilon^{\rm{CH}}_{i~\rm{PM}}}
\end{eqnarray}
where $N^{\rm{sel}}$ is the number of selected events, $\Phi$ is the integrated \nmb (\nmbm) flux, $T$ is the number of target nucleons, and $\varepsilon$ is the signal-selection efficiency. The $N^{\rm{BG}}$ is the number of expected backgrounds, and $N^{\rm{BG}}_{\rm{WM}}$ is estimated not only by the MC simulation but also by the calculated cross-section on the hydrocarbon target to take into account the contribution from the plastic scintillators in the WAGASCI module. Quantities, $\Phi$ and $T$, are summarized in Table~\ref{tab:quantities}. The $U_{ij}$ is an unfolding matrix which is iteratively calculated based on the D'Agostini method~\cite{cite:unfolding}. To avoid any dependence of $U_{ij}$ on the input neutrino interaction simulation the number of iterations is not truncated but rather ran through to convergence such that the result is effectively unregularized (more details are presented in Sec.~\ref{sec:result}). In the unfolding procedure, we choose a flat prior, and define the number of iterations as 1500. The subscripts of WM and PM represent the WAGASCI module and the Proton Module, respectively, and those of $\rm H_{2}O$ and CH represent target materials. 

\begin{table}[!h]
	\centering
	\caption{Summary of integrated neutrino fluxes and the number of target nucleons used for cross-section calculation. The fluxes are normalized to the actual recorded POT ($7.91\times10^{20}$).}
	\label{tab:quantities}
	\begin{tabular}{ccccc}
    \hline\hline
    Cross section                & $\Phi_{\overline{\nu}_{\mu}}$                  & $\Phi_{\nu_{\mu}}$                         & $T_{\rm H_{2}O}$          & $T_{\rm CH}$ \\
    \hline
    $\sigma_{\rm H_{2}O}$     & $1.69 \times 10^{13}~{\rm cm}^{-2}$  & $1.48 \times 10^{12}~{\rm cm}^{-2}$  & $4.957 \times 10^{28}$  & $1.107 \times 10^{28}$  \\
    $\sigma_{\rm CH}$         & $1.70 \times 10^{13}~{\rm cm}^{-2}$  & $1.49 \times 10^{12}~{\rm cm}^{-2}$  &  -                               & $9.210 \times 10^{28}$ \\
    \hline\hline
    \end{tabular}
\end{table}

%%% Backgrounds
All of the backgrounds are estimated by the MC simulation, except for interactions on WAGASCI plastic scintillators. They constitute one of main background sources for \sigho, since about 20\% of the fiducial volume of the WAGASCI module is occupied by plastic scintillators. They are calculated by normalizing from the number of selected events in the Proton Module.

%%%%%%%%%%%%%%%%%%%%%%%%%%%%
%%% Cross-section extraction numubar+numu %%%
%%%%%%%%%%%%%%%%%%%%%%%%%%%%
\subsection{\nmb cross-sections and \nmbm cross-sections}\label{sec:numu+numubar}
As shown in Tables~\ref{tab:event_summary_WM} and \ref{tab:event_summary_PM}, the \nm CC interactions are the dominant background and are irreducible in our \nmb event selection since we cannot determine the charge of the outgoing muon. In order to be less model-dependent, we also measured a combined \nmbm cross-section, since this measurement does not rely on model assumptions to subtract the \nm background.\\
\indent The event selection, the number of selected events ($N_{j}^{\rm sel}$), and the number of target nucleons ($T$) are common to the \nmb and \nmbm cross-section measurements. Differences between these measurements are summarized in Table~\ref{tab:difference}.

\begin{table}[!h]
	\centering
	\caption{Summary of differences between the \nmb and \nmbm cross-section measurements.}
	\label{tab:difference}
	\begin{tabular}{l|c|c}
    \hline
    \hline
                         & \nmb cross-section   & \nmbm cross-section  \\
    \hline
    ~~\nm CC interaction  & Background                      & Signal \\
    \hline
    \begin{tabular}{c}Detection efficiency ($\varepsilon$)\\ Unfolding matrix ($U_{ij})$ \end{tabular} & Calculated with \nmb samples & Calculated with \nmbm samples \\
    \hline
    ~~Integrated flux            & $\Phi_{\overline{\nu}_{\mu}}$        & $\Phi_{\overline{\nu}_{\mu} + \nu_{\mu}}$ \\
    \hline
    \hline
    \end{tabular}
\end{table}

% SECTION: SYSTEMATIC UNCERTAINTIES %%%
\section{Uncertainties}\label{sec:uncertainty}
Evaluation methods for each uncertainty are almost the same as those considered in Ref.~\cite{cite:koga}, and detailed information can be found in that reference.
%%%%%%%%%%%%%%%%%%%%%%
%%% Flux uncertainties %%%
%%%%%%%%%%%%%%%%%%%%%%
\subsection{Systematic uncertainties from neutrino flux uncertainties}\label{sec:sys_flux}
The uncertainty on the neutrino flux is estimated according to knowledge of hadron interactions and the J-PARC beamline.
For systematic uncertainties on the cross-section extraction, effects on the number of background events ($N^{\rm BG}$), integrated flux ($\Phi$), detection efficiency ($\varepsilon$), and the unfolding matrix ($U_{ij}$) are considered. Events generated in the MC simulation are varied based on the estimated flux uncertainties in bins of the true neutrino energies, and correlations among them. The variation of the cross-section is calculated by using 10,000 toy samples accordingly. The $\pm1\sigma$ range of the distribution is taken as the systematic uncertainty.

% Ratio
The uncertainties from neutrino flux on the integrated cross-section for \ccnoppf~are expected to be about 10\% for \sigho~and \sigch, and they give the dominant contributions to the total uncertainty. On the other hand, the uncertainties for the cross-section ratio (\sigratio) are about 0.5\%, since most of the parameters are strongly correlated and the uncertainties cancel.

%%%%%%%%%%%%%%%%%%%%%%%%%%%%%%
%%% Neutrino interaction model %%%
%%%%%%%%%%%%%%%%%%%%%%%%%%%%%%

\subsection{Systematic uncertainties from neutrino-interaction model } \label{sec:sys_neut}

%%% NEUT Paramters
Uncertainties on the neutrino-interaction model are estimated based on the understanding of the model applied to the MC-event generator. Each parameter related to this analysis is varied to cover model uncertainties, and the propagation to the extracted cross-sections is calculated. The parameters with their default values and 1$\sigma$ variations are summarized in Table~\ref{tab:neut_param}. When the uncertainty is calculated, no correlation is assumed between different target nuclei for the Fermi momentum ($\mathrm{P_{f}}$), binding energy ($\mathrm{E_{b}}$), 2p2h, CC coherent parameters, and nucleon final state interactions (FSI). Full correlation between the different targets is assumed for the other parameters.

\begin{table}[htb]
\centering
\caption{Summary of the default values of the parameters used in the neutrino-interaction model and their uncertainties.}
\label{tab:neut_param}
\begin{tabular}{lcc}
\hline
\hline
Parameter                                   & Nominal value & Uncertainties (1$\sigma$)    \\

\hline
\hline
CCQE: the Smith-Moniz model & & \\
RFG with the RPA correction & & \\\cline{1-1}
\\[-1em]
$\rm M^{QE}_{A}$                            & 1.15~GeV      & 0.18~GeV              \\
$\rm P_f~^{12}C$                            & 223~MeV       & 31~MeV               \\
$\rm P_f~^{16}O$                            & 225~MeV       & 31~MeV               \\
$\rm E_b~^{12}C$                            & 25~MeV        & 9~MeV                \\
$\rm E_b~^{16}O$                            & 27~MeV        & 9~MeV                \\

\hline
\hline
2p2h: the Nieves model & & \\
\cline{1-1}
$\rm 2p2h~normalization~^{12}C$             & 100~\%        & 100~\%                \\
$\rm 2p2h~normalization~^{16}O$             & 100~\%        & 100~\%                \\
$\rm 2p2h~shape~^{12}C$                     & 100~\%        & 100~\%                \\
$\rm 2p2h~shape~^{16}O$                     & 100~\%        & 100~\%                \\

\hline
\hline
CC-resonant model: the Rein-Sehgal model \\
\cline{1-1}
$\rm M^{Res}_{A}$                           & 0.95~GeV      & 0.15~GeV                 \\
$\rm C_{A5}$                                & 1.01          & 0.12                 \\
$\rm Isospin~ \frac{1}{2}~ background$                & 1.30          & 0.20                 \\

\hline
\hline
CC coherent model: the Berger-Sehgal model & & \\
\cline{1-1}
$\rm CCcoh~normalization~^{12}C$            & 100~\%        & 100~\%               \\
$\rm CCcoh~normalization~^{16}O$            & 100~\%        & 100~\%               \\

\hline
\hline
DIS: GRV98 PDF with Bodek-Yang modifications & & \\
\cline{1-1}
$\rm DIS~ correction~ factor$                         &  $x=0$             & $x=0.40$                 \\

\hline
\hline
NC interactions & & \\
\cline{1-1}
$\rm NCcoh~norm$                            & 100~\%        & 30~\%                \\
$\rm NCother~norm$                          & 100~\%        & 30~\%                \\

\hline
\hline
Final state interactions of pions & & \\
\cline{1-1}
Pion Absorption normalization               & 1.1        & 50~\%                \\
Pion Charge Exchange (low E) normalization  & 1.0        & 50~\%                \\
Pion Charge Exchange (high E) normalization & 1.8        & 30~\%                \\
Pion Quasi Elastic (low E) normalization    & 1.0        & 50~\%                \\
Pion Quasi Elastic (high E) normalization   & 1.8        & 30~\%                \\
Pion Inelastic normalization                & 1.0        & 50~\%                \\

\hline
\hline
Final state interactions of nucleons & & \\
\cline{1-1}
Nucleon FSI               & 100~\%        & 100~\%                \\
\hline
\hline
\end{tabular}
\end{table}

%% Summary
The uncertainties due to the neutrino-interaction model are dominated by effects from CCQE and 2p2h interactions, and nucleon FSI, followed by pion production ($\rm M_A^{Res}$ and $\rm C_{A5}$) and Fermi momentum ($\mathrm{P_{f}}$). The CCQE and 2p2h interactions have uncertainties that are 2\% larger than other categories and have the largest effect on the detection-efficiency estimation, since they dominate the \ccnopp~signal and then largely distort the prior distribution. Nucleon FSI mainly affect the number of backgrounds via \nm interactions, since more nucleons often exist in the final state of \nm interactions than that of \nmb interactions. Hence, this effect becomes smaller for the \nmbm cross-section measurement.

%%%%%%%%%%%%%%%%%%%%%%%
%%% Detector response %%%
%%%%%%%%%%%%%%%%%%%%%%%
\subsection{Systematic uncertainties from detector response } \label{ sec:sys_det }

%%%% Overall
Uncertainties on the detector response are estimated based on measurements during the detector construction, commissioning data taking with cosmic muons, and operation with the anti-neutrino beam. Effects on the number of selected events are estimated according to the uncertainty on the detector response, and the systematic uncertainty on the cross-section measurement is estimated by applying fluctuations to the measured number of selected events. In order to apply fluctuations to the number of selected events, no correlation between \WP~is assumed, except for the beam-related backgrounds which should be common between the two detectors. Correlations between each bin of reconstructed tracks are considered. The target mass, MPPC noise, scintillator crosstalk, reconstruction efficiency, event pileup, beam-related backgrounds, and event selections are considered as sources of uncertainty. Uncertainties from the event selection are estimated from the difference between data and simulation in the variation of the number of selected events for each selection criterion.

\subsection{Total uncertainty}
Total uncertainties are summarized in Appendix~\ref{appendix:xsec}. For the cross-section measurements of \ccnoppf, the total uncertainty on the absolute cross-sections, \sigho~and \sigch, is dominated by the neutrino-flux uncertainty, while that on the cross-section ratio, \sigratio, is dominated by statistical errors and errors on the detector response.

% SECTION: RESULTS %%%
\section{Results}\label{sec:result}
\begin{figure}[tbh]
\centering
\includegraphics[width=\linewidth]{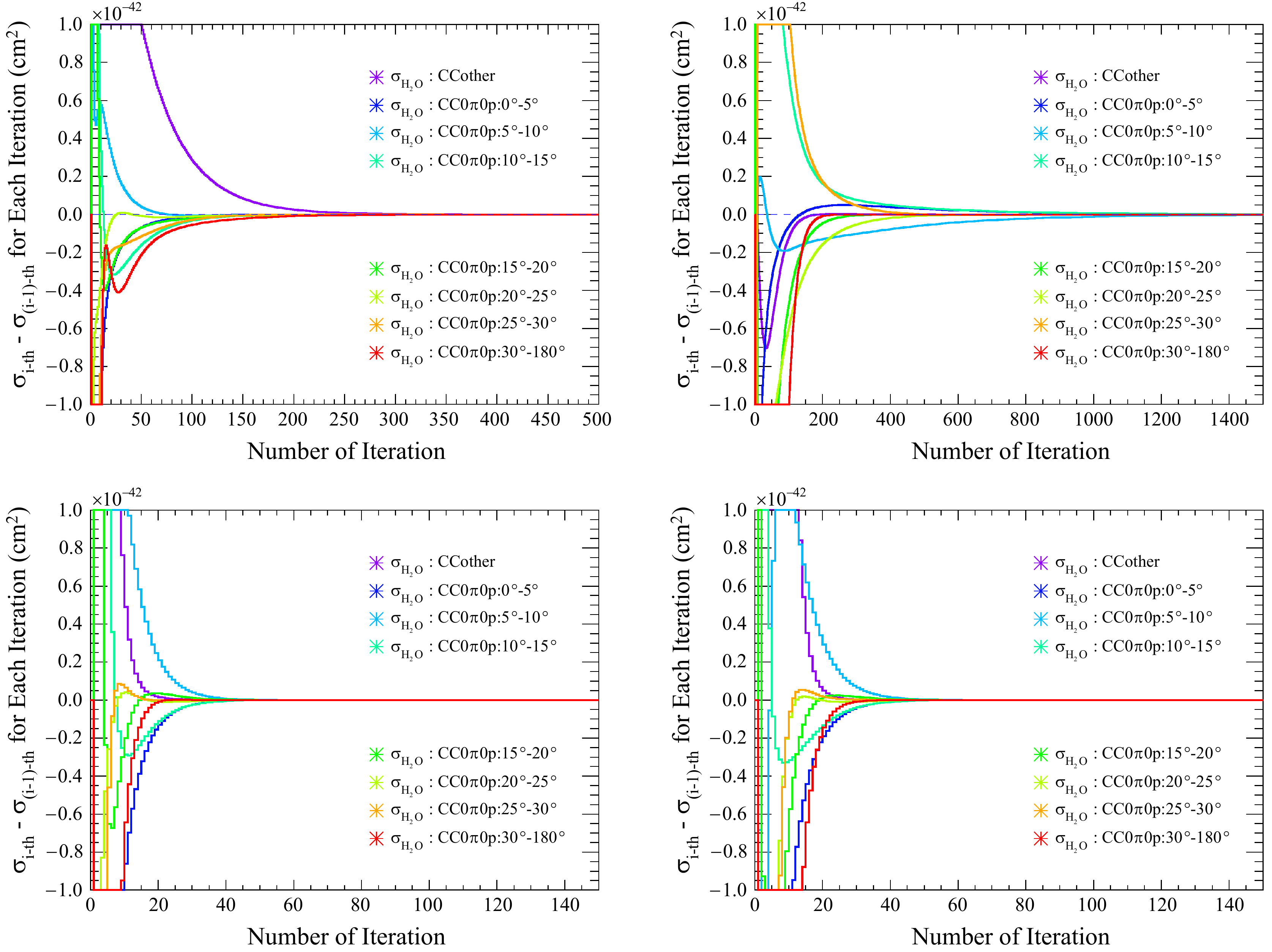}
\caption{Convergence of the extracted cross-sections, \nmb cross-section (left) and \nmbm cross-section (right). Top line: $\sigma_{\rm H_2O}$. Bottom line: $\sigma_{\rm CH}$. The plots show the first 500 iterations for \nmb $\rm \sigma_{H_2O}$, 150 iterations for \nmb $\rm \sigma_{CH}$, 1500 iterations for \nmbm $\rm \sigma_{H_2O}$, 150 iterations for \nmbm $\rm \sigma_{CH}$.}
\label{fig:result_convergence2}
\end{figure}

%%% Convergence 
%%% 
Figure~\ref{fig:result_convergence2} shows the convergence of extracted cross-sections with respect to the number of iterations. %Here, drifts of the extracted cross-sections, which is defined as $\sigma_{i\mathchar`-{\rm th~iteration}} - \sigma_{(i-1)\mathchar`-{\rm th~iteration}}$, are shown.
Each cross-section converges to a constant value after 1500 iterations.

%%% Measured values

\begin{figure}[!h]
\centering
\includegraphics[width=\linewidth]{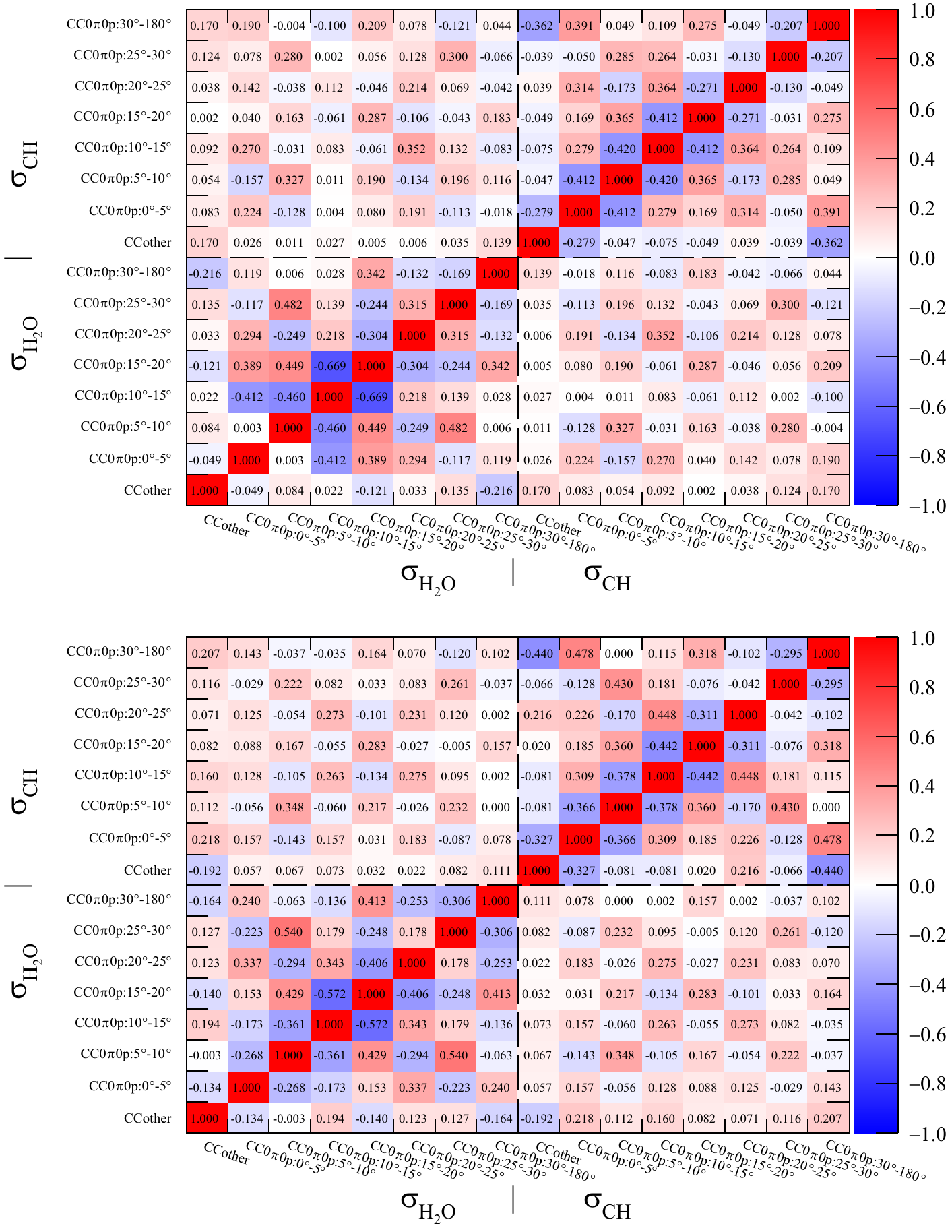}
\caption{Correlation matrices including all uncertainties for \nmb (top) and \nmbm (bottom) cross sections.}
\label{fig:corr_matrix}
\end{figure}

\begin{figure}[!h]
\centering
\includegraphics[width=\linewidth]{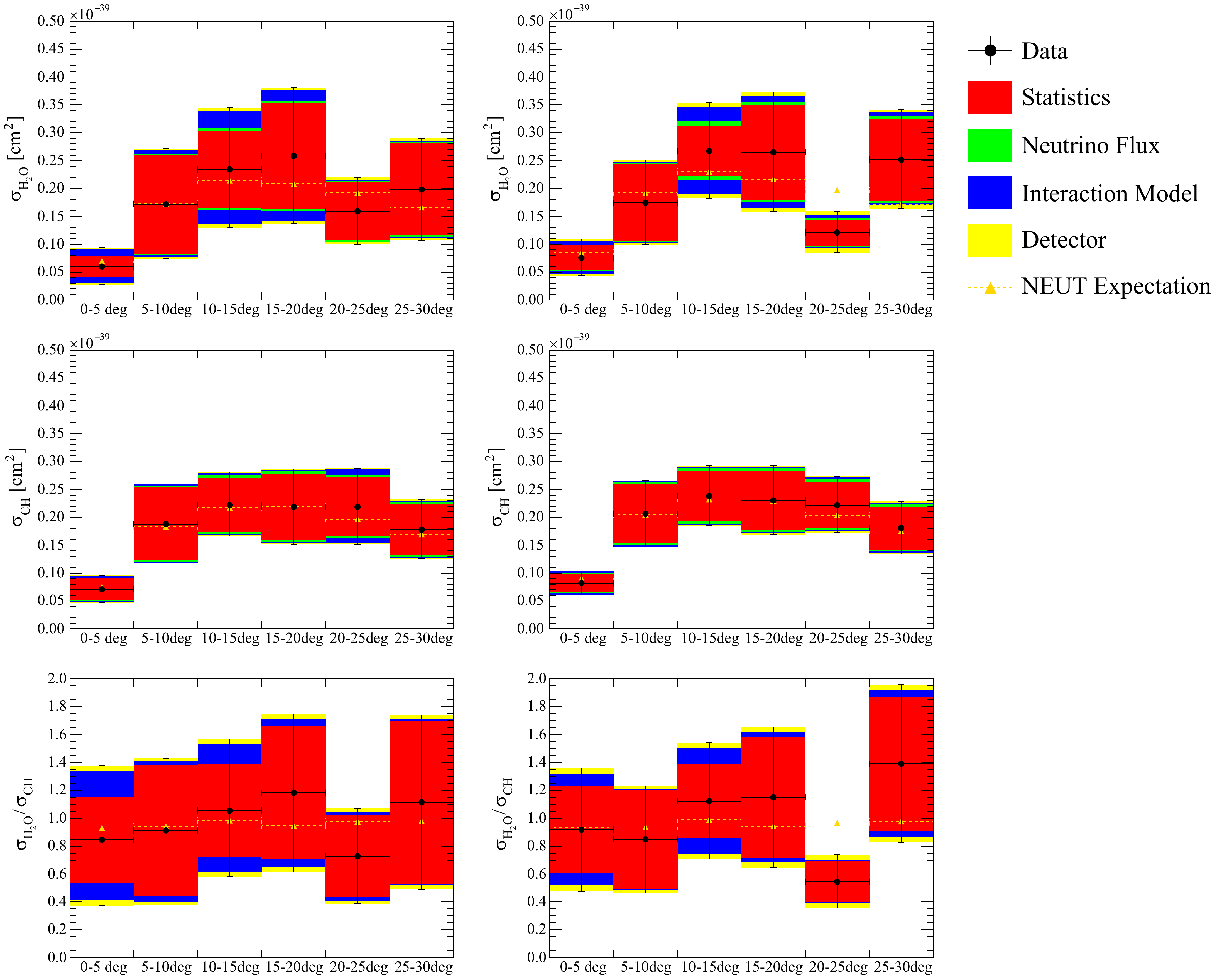}
\caption{Measured values for the differential \nmb cross-section (left) and \nmbm cross-section (right). Top line: $\sigma_{\rm H_2O}$. Middle line: $\sigma_{\rm CH}$. Bottom line: $\sigma_{\rm H_2O}/\sigma_{\rm CH}$ (bottom). Each plot shows the cumulative quadratic sum of the uncertainties from statistics, neutrino flux, neutrino-interaction model and detector response.}
\label{fig:result_differential}
\end{figure}

The measured flux-integrated CC0$\pi$0p cross-sections on $\rm H_{2}O$ and CH are summarized as follows:
\begin{eqnarray}
	\sigma_{\rm H_2O}^{\overline{\nu}_\mu} &=&  \left[ 1.082 \pm 0.068(\rm stat.)^{+0.145}_{-0.128}(\rm syst.) \right] \times10^{-39} {\rm  cm^2/ nucleon}, \nonumber \\
	\sigma_{\rm CH}^{\overline{\nu}_\mu} &=&  \left[ 1.096 \pm 0.054(\rm stat.)^{+0.132}_{-0.117}(\rm syst.) \right] \times10^{-39} {\rm cm^2 / nucleon }, \nonumber \\
	\sigma_{\rm H_2O}^{\overline{\nu}_\mu}/\sigma_{\rm CH}^{\overline{\nu}_\mu} &=&  0.987 \pm 0.078(\rm stat.)^{+0.093}_{-0.090}(\rm syst.), \nonumber \\
	\sigma_{\rm H_2O}^{\overline{\nu}_\mu+\nu_\mu} &=&  \left[ 1.155 \pm 0.064(\rm stat.)^{+0.148}_{-0.129}(\rm syst.) \right] \times10^{-39} {\rm cm^2 / nucleon }, \nonumber \\
	\sigma_{\rm CH}^{\overline{\nu}_\mu+\nu_\mu} &=&  \left[ 1.159 \pm 0.049(\rm stat.)^{+0.129}_{-0.115}(\rm syst.) \right] \times10^{-39} {\rm cm^2 / nucleon }, \nonumber \\
	\sigma_{\rm H_2O}^{\overline{\nu}_\mu+\nu_\mu}/\sigma_{\rm CH}^{\overline{\nu}_\mu+\nu_\mu} &=&  0.996 \pm 0.069(\rm stat.)^{+0.083}_{-0.078}(\rm syst.), \nonumber
\end{eqnarray}
where the cross-sections are normalized by the number of all nucleons in molecules of \hto~and CH.
All the integrated cross-sections are consistent with the models in the MC-event generator, NEUT, within a level of 1$\sigma$. Figure~\ref{fig:corr_matrix} shows correlation matrices including all uncertainties for \nmb (top) and \nmbm (bottom) cross sections.
Figure~\ref{fig:result_differential} shows the distributions of the measured differential cross-sections for \ccnoppf, with their uncertainties and expectations from NEUT (5.3.3). Basically, the measured cross-sections on each phase-space bin agree with the NEUT expectation within 1$\sigma$, except for $\sigma_{\rm H_2O}^{\overline{\nu}_\mu+\nu_\mu}$ and $\sigma_{\rm H_2O}^{\overline{\nu}_\mu+\nu_\mu}/\sigma_{\rm CH}^{\overline{\nu}_\mu+\nu_\mu}$ in the phase-space region of $20\mathchar`-25^\circ$.

%% Comparison with  predictions:  Chi2
In order to evaluate the agreement of measured differential cross-sections with predictions, $\chi^2$ values are calculated based on the total uncertainty including both the statistical and systematic uncertainties. Tables~\ref{tab:result_chi1} and \ref{tab:result_chi2} show the calculated $\chi^2$ values for the predictions from NEUT and GENIE. Considering the number of degrees of freedom is eight, the calculated $\chi^2$ values suggest that the measured cross-sections agree well with neutrino-interaction models implemented in those generators.

%%%%%%%%%%%%%%%%%%%%%%%%%%%%%%%%%%%%%%%%%%%%%%%%%%%%

\begin{table}[!h]
\centering
\caption{Absolute $\chi^2$ values for the \nmb and \nmbm cross-sections, with respect to the total uncertainty.}
\label{tab:result_chi1}
\begin{tabular}{|l|ccc|ccc|}
\hline
\hline
 & \multicolumn{3}{c|}{\nmb~cross section} & \multicolumn{3}{c|}{\nmbm~cross section} \\
\cline{2-7}
 & $\sigma_{H_2O}$ & $\sigma_{CH}$ & $\sigma_{H_2O}$/$\sigma_{CH}$
 & $\sigma_{H_2O}$ & $\sigma_{CH}$ & $\sigma_{H_2O}$/$\sigma_{CH}$ \\
\hline
\hline
NEUT  & 3.19 & 11.34 & 1.71 & 7.06 & 2.63 & 6.87 \\
GENIE & 4.25 & 14.26 & 1.83 & 7.09 & 3.38 & 7.55 \\
\hline
\hline
\end{tabular}
\end{table}

\begin{table}[!h]
\centering
\caption{Absolute $\chi^2$ values for the \nmb and \nmbm cross-sections only for a muon
angle less than 30 degrees, concerning the total uncertainty. }
\label{tab:result_chi2}
\begin{tabular}{|l|ccc|ccc|}
\hline
\hline
 & \multicolumn{3}{c|}{\nmb~cross section} & \multicolumn{3}{c|}{\nmbm~cross section} \\
\cline{2-7}
 & $\sigma_{H_2O}$ & $\sigma_{CH}$ & $\sigma_{H_2O}$/$\sigma_{CH}$
 & $\sigma_{H_2O}$ & $\sigma_{CH}$ & $\sigma_{H_2O}$/$\sigma_{CH}$ \\
\hline
\hline
NEUT  & 0.74 & 0.16 & 0.81 & 5.93 & 0.33 & 5.76 \\
GENIE & 0.72 & 0.54 & 0.89 & 5.98 & 0.57 & 6.35 \\
\hline
\hline
\end{tabular}
\end{table}

% SECTION: CONCLUSION %%%
\section{Conclusion}\label{sec:conclusion}
%%%%%%%%%%%%%%%%%%
%%% Conclusion %%%
%%%%%%%%%%%%%%%%%%
In this paper, we report measurements of (anti-)neutrino cross-sections on water and hydrocarbon targets with the WAGASCI module and the Proton Module using the T2K anti-neutrino beam. The mean neutrino energy is 0.86~GeV, and the peak is at 0.66~GeV with 1$\sigma$ spread of $^{+0.40}_{-0.25}$~GeV. The signal is taken to be charged-current interactions with one muon and no pions nor protons produced, based on the kinematic measurements of muons, pions, and protons. The differential cross-sections and integrated cross-sections for the \nmb only and \nmbm fluxes are measured. The results agree with the current neutrino-interaction models used in the T2K oscillation analysis within their statistical and systematic uncertainties.

% Acknowledgements
\section*{Acknowledgements}
We thank the J-PARC staff for the superb accelerator performance. We thank the CERN NA61/SHINE Collaboration for providing valuable particle production data. We acknowledge the support of MEXT, Japan; NSERC (Grant No. SAPPJ-2014-00031), NRC and CFI, Canada; CEA and CNRS/IN2P3, France; DFG, Germany; INFN, Italy; National Science Centre (NCN) and Ministry of Science and Higher Education, Poland; RSF (Grant \#19-12-00325) and Ministry of Science and Higher Education, Russia; MINECO and ERDF funds, Spain; SNSF and SERI, Switzerland; STFC, UK; and DOE, USA. We also thank CERN for the UA1/NOMAD magnet, DESY for the HERA-B magnet mover system, NII for SINET4, the WestGrid and SciNet consortia in Compute Canada, and GridPP in the United Kingdom. In addition, participation of individual researchers and institutions has been further supported by funds from ERC (FP7), "la Caixa” Foundation (ID 100010434, fellowship code LCF/BQ/IN17/11620050), the European Union’s Horizon 2020 Research and Innovation programme under the Marie Sklodowska-Curie grant agreement no. 713673 and H2020 Grant No. RISE-GA644294-JENNIFER 2020; JSPS, Japan; Royal Society, UK; and the DOE Early Career program, USA.

% Appendix
\appendix
\section{Appendix: Differential Cross-Sections}\label{appendix:xsec}
The \nmb and \nmbm cross-section measurements are summarized in Tables~\ref{tab:measurement1} and \ref{tab:measurement2}. Total uncertainties are summarized in Tables~\ref{tab:totalerror1} and \ref{tab:totalerror2}. The total uncertainty on the right-hand column is calculated as a quadratic sum of the statistical uncertainty and the systematic uncertainties. By taking the water to hydrocarbon cross-section ratio, uncertainties on the T2K (anti-)neutrino beam prediction, which are the dominant errors for absolute cross-section measurements, largely cancel.

\begin{table}[!h]
\centering
\caption{Summary of the \nmb cross-section measurement. Units for \sigho~and \sigch~are $[\rm \times10^{-39}~cm^2 / nucleon]$.}
\label{tab:measurement1}
\begin{tabular}{llccc}
\hline
\hline
                                    & True phase space         & Cross section & Stat. err. & Syst. err. \\
\hline
\hline
$\rm \sigma_{H_{2}O}              $
                                    & $\rm CCother$            & 0.244         & $\pm0.120$      & $+0.206/-0.199$ \\
                                    & $\rm CC0\pi0p:0$-$5^\circ$    & 0.060         & $\pm0.018$      & $+0.029/-0.027$ \\
                                    & $\rm CC0\pi0p:5$-$10^\circ$   & 0.172         & $\pm0.089$      & $+0.045/-0.040$ \\
                                    & $\rm CC0\pi0p:10$-$15^\circ$  & 0.234         & $\pm0.069$      & $+0.086/-0.080$ \\
                                    & $\rm CC0\pi0p:15$-$20^\circ$  & 0.259         & $\pm0.095$      & $+0.077/-0.075$ \\
                                    & $\rm CC0\pi0p:20$-$25^\circ$  & 0.159         & $\pm0.052$      & $+0.031/-0.029$ \\
                                    & $\rm CC0\pi0p:25$-$30^\circ$  & 0.198         & $\pm0.082$      & $+0.040/-0.039$ \\
                                    & $\rm CC0\pi0p:30$-$180^\circ$ & 0.635         & $\pm0.145$      & $+0.243/-0.217$ \\
\hline
                                    & $\rm CC Total$           & 1.961         & $\pm0.196$      & $+0.400/-0.368$ \\
                                    & $\rm CC0\pi0p:0$-$30^\circ$   & 1.082         & $\pm0.068$      & $+0.145/-0.128$ \\
\hline
\hline
$\rm \sigma_{CH}                  $
                                    & $\rm CCother$            & 0.162         & $\pm0.057$      & $+0.149/-0.141$ \\
                                    & $\rm CC0\pi0p:0$-$5^\circ$    & 0.071         & $\pm0.019$      & $+0.015/-0.014$ \\
                                    & $\rm CC0\pi0p:5$-$10^\circ$   & 0.188         & $\pm0.065$      & $+0.029/-0.025$ \\
                                    & $\rm CC0\pi0p:10$-$15^\circ$  & 0.222         & $\pm0.048$      & $+0.034/-0.028$ \\
                                    & $\rm CC0\pi0p:15$-$20^\circ$  & 0.219         & $\pm0.060$      & $+0.033/-0.030$ \\
                                    & $\rm CC0\pi0p:20$-$25^\circ$  & 0.219         & $\pm0.052$      & $+0.045/-0.042$ \\
                                    & $\rm CC0\pi0p:25$-$30^\circ$  & 0.178         & $\pm0.046$      & $+0.028/-0.025$ \\
                                    & $\rm CC0\pi0p:30$-$180^\circ$ & 0.975         & $\pm0.285$      & $+0.320/-0.344$ \\
\hline
                                    & $\rm CC Total$           & 2.233         & $\pm0.195$      & $+0.498/-0.446$ \\
                                    & $\rm CC0\pi0p:0$-$30^\circ$   & 1.096         & $\pm0.054$      & $+0.132/-0.117$ \\
\hline
\hline
$\rm \sigma_{H_{2}O}/\sigma_{CH} $
                                    & $\rm CCother$            & 1.508         & $\pm0.753$      & $+1.803/-1.524$ \\
                                    & $\rm CC0\pi0p:0$-$5^\circ$    & 0.846         & $\pm0.310$      & $+0.431/-0.356$ \\
                                    & $\rm CC0\pi0p:5$-$10^\circ$   & 0.913         & $\pm0.470$      & $+0.213/-0.253$ \\
                                    & $\rm CC0\pi0p:10$-$15^\circ$  & 1.056         & $\pm0.334$      & $+0.389/-0.336$ \\
                                    & $\rm CC0\pi0p:15$-$20^\circ$  & 1.183         & $\pm0.476$      & $+0.306/-0.312$ \\
                                    & $\rm CC0\pi0p:20$-$25^\circ$  & 0.728         & $\pm0.291$      & $+0.176/-0.179$ \\
                                    & $\rm CC0\pi0p:25$-$30^\circ$  & 1.115         & $\pm0.583$      & $+0.227/-0.219$ \\
                                    & $\rm CC0\pi0p:30$-$180^\circ$ & 0.651         & $\pm0.241$      & $+0.291/-0.299$ \\
\hline
                                    & $\rm CC Total$           & 0.878         & $\pm0.111$      & $+0.177/-0.191$ \\
                                    & $\rm CC0\pi0p:0$-$30^\circ$   & 0.987         & $\pm0.078$      & $+0.093/-0.090$ \\
\hline
\hline
\end{tabular}

\end{table}

\begin{table}[!h]
\centering
\caption{Summary of the \nmbm cross-section measurement. Units for \sigho~and \sigch~are $[\rm \times10^{-39}~cm^2 / nucleon]$.}
\label{tab:measurement2}
\begin{tabular}{llccc}
\hline
\hline
                                    & True phase space         & Cross section & Stat. err. & Syst. err. \\
\hline
\hline
$\rm \sigma_{H_{2}O}              $
                                    & $\rm CCother$                      & 0.923         & $\pm0.126$      & $+0.224/-0.219$ \\
                                    & $\rm CC0\pi0p:0$-$5^\circ$    & 0.075         & $\pm0.022$      & $+0.026/-0.023$ \\
                                    & $\rm CC0\pi0p:5$-$10^\circ$   & 0.175         & $\pm0.069$      & $+0.034/-0.032$ \\
                                    & $\rm CC0\pi0p:10$-$15^\circ$  & 0.267         & $\pm0.045$      & $+0.074/-0.072$ \\
                                    & $\rm CC0\pi0p:15$-$20^\circ$  & 0.265         & $\pm0.085$      & $+0.067/-0.065$ \\
                                    & $\rm CC0\pi0p:20$-$25^\circ$  & 0.121         & $\pm0.023$      & $+0.030/-0.027$ \\
                                    & $\rm CC0\pi0p:25$-$30^\circ$  & 0.252         & $\pm0.074$      & $+0.051/-0.046$ \\
                                    & $\rm CC0\pi0p:30$-$180^\circ$ & 0.590         & $\pm0.185$      & $+0.235/-0.217$ \\
\hline
                                    & $\rm CC Total$                       & 2.668         & $\pm0.171$      & $+0.353/-0.327$ \\
                                    & $\rm CC0\pi0p:0$-$30^\circ$    & 1.155         & $\pm0.064$      & $+0.148/-0.129$ \\
\hline
\hline
$\rm \sigma_{CH}                  $
                                    & $\rm CCother$                      & 0.877         & $\pm0.062$      & $+0.364/-0.344$ \\
                                    & $\rm CC0\pi0p:0$-$5^\circ$    & 0.082         & $\pm0.016$      & $+0.015/-0.014$ \\
                                    & $\rm CC0\pi0p:5$-$10^\circ$   & 0.206         & $\pm0.053$      & $+0.028/-0.025$ \\
                                    & $\rm CC0\pi0p:10$-$15^\circ$  & 0.238         & $\pm0.045$      & $+0.030/-0.027$ \\
                                    & $\rm CC0\pi0p:15$-$20^\circ$  & 0.230         & $\pm0.053$      & $+0.033/-0.030$ \\
                                    & $\rm CC0\pi0p:20$-$25^\circ$  & 0.222         & $\pm0.040$      & $+0.032/-0.029$ \\
                                    & $\rm CC0\pi0p:25$-$30^\circ$  & 0.181         & $\pm0.038$      & $+0.028/-0.026$ \\
                                    & $\rm CC0\pi0p:30$-$180^\circ$ & 0.969         & $\pm0.228$      & $+0.280/-0.309$ \\
\hline
                                    & $\rm CC Total$                       & 3.005         & $\pm0.175$      & $+0.499/-0.444$ \\
                                    & $\rm CC0\pi0p:0$-$30^\circ$    & 1.159         & $\pm0.049$      & $+0.129/-0.115$ \\
\hline
\hline
$\rm \sigma_{H_{2}O}/\sigma_{CH} $
                                    & $\rm CCother$                       & 1.052         & $\pm0.169$      & $+0.563/-0.519$ \\
                                    & $\rm CC0\pi0p:0$-$5^\circ$    & 0.919         & $\pm0.310$      & $+0.318/-0.316$ \\
                                    & $\rm CC0\pi0p:5$-$10^\circ$   & 0.848         & $\pm0.352$      & $+0.152/-0.156$ \\
                                    & $\rm CC0\pi0p:10$-$15^\circ$  & 1.123         & $\pm0.265$      & $+0.325/-0.321$ \\
                                    & $\rm CC0\pi0p:15$-$20^\circ$  & 1.151         & $\pm0.435$      & $+0.256/-0.255$ \\
                                    & $\rm CC0\pi0p:20$-$25^\circ$  & 0.546         & $\pm0.143$      & $+0.127/-0.124$ \\
                                    & $\rm CC0\pi0p:25$-$30^\circ$  & 1.391         & $\pm0.482$      & $+0.300/-0.293$ \\
                                    & $\rm CC0\pi0p:30$-$180^\circ$ & 0.609         & $\pm0.251$      & $+0.297/-0.273$ \\
\hline
                                    & $\rm CC Total$                       & 0.888         & $\pm0.077$      & $+0.148/-0.160$ \\
                                    & $\rm CC0\pi0p:0$-$30^\circ$    & 0.996         & $\pm0.069$      & $+0.083/-0.078$ \\
\hline
\hline
\end{tabular}

\end{table}

\begin{table}[!h]
\centering
\caption{Summary of systematic uncertainties on the \nmb cross-section [\%].}
\label{tab:totalerror1}
\scalebox{0.9}{
\begin{tabular}{clccccc} 
\hline
\hline
 & True phase space  & Statistics & Neutrino & Neutrino     & Detector & Total \\
 &                   &            &   flux   & interactions & response &       \\
\hline
\hline
$\rm \sigma_{H_{2}O}$
 &  $\rm CCother$                  & $\pm49.4$ & $^{+8.5}_{-7.2}$ & $^{+57.3}_{-53.2}$ & $\pm61.7$ & $^{+98.0}_{-95.5}$  \\
 &  $\rm CC0\pi0p:0$-$5^\circ$     & $\pm29.5$ & $^{+10.7}_{-9.0}$ & $^{+41.2}_{-37.7}$ & $\pm22.2$ & $^{+56.4}_{-53.5}$  \\
 &  $\rm CC0\pi0p:5$-$10^\circ$    & $\pm51.6$ & $^{+11.4}_{-9.5}$ & $^{+19.4}_{-16.1}$ & $\pm13.9$ & $^{+58.0}_{-56.6}$  \\
 &  $\rm CC0\pi0p:10$-$15^\circ$   & $\pm29.3$ & $^{+11.0}_{-9.3}$ & $^{+31.6}_{-28.8}$ & $\pm15.3$ & $^{+47.1}_{-44.9}$  \\
 &  $\rm CC0\pi0p:15$-$20^\circ$   & $\pm36.7$ & $^{+10.9}_{-9.3}$ & $^{+24.3}_{-24.0}$ & $\pm13.3$ & $^{+47.3}_{-46.7}$  \\
 &  $\rm CC0\pi0p:20$-$25^\circ$   & $\pm32.7$ & $^{+10.6}_{-9.1}$ & $^{+8.3}_{-6.8}$ & $\pm14.2$ & $^{+38.1}_{-37.4}$  \\
 &  $\rm CC0\pi0p:25$-$30^\circ$   & $\pm41.5$ & $^{+11.1}_{-9.4}$ & $^{+8.2}_{-8.9}$ & $\pm14.9$ & $^{+46.2}_{-45.9}$  \\
 &  $\rm CC0\pi0p:30$-$180^\circ$  & $\pm22.8$ & $^{+10.3}_{-8.7}$ & $^{+29.4}_{-24.5}$ & $\pm22.2$ & $^{+44.5}_{-41.1}$  \\
\hline
 &  $\rm CC Total$                 & $\pm10.0$ & $^{+9.6}_{-8.2}$ & $^{+15.7}_{-14.3}$ & $\pm9.0$ & $^{+22.7}_{-21.3}$  \\
 &  $\rm CC0\pi0p:0$-$30^\circ$    & $\pm6.3$ & $^{+10.9}_{-9.2}$ & $^{+5.6}_{-5.0}$ & $\pm5.5$ & $^{+14.8}_{-13.4}$  \\
\hline
\hline
$\rm \sigma_{CH}$
 &  $\rm CCother$                  & $\pm35.5$ & $^{+8.6}_{-7.4}$ & $^{+81.8}_{-76.1}$ & $\pm41.2$ & $^{+98.6}_{-93.8}$  \\
 &  $\rm CC0\pi0p:0$-$5^\circ$     & $\pm27.1$ & $^{+10.8}_{-9.2}$ & $^{+17.3}_{-16.4}$ & $\pm7.4$ & $^{+34.8}_{-33.9}$  \\
 &  $\rm CC0\pi0p:5$-$10^\circ$    & $\pm34.8$ & $^{+10.8}_{-9.1}$ & $^{+9.6}_{-7.8}$ & $\pm5.8$ & $^{+38.1}_{-37.2}$  \\
 &  $\rm CC0\pi0p:10$-$15^\circ$   & $\pm21.7$ & $^{+10.4}_{-9.0}$ & $^{+9.3}_{-5.9}$ & $\pm6.6$ & $^{+26.6}_{-25.1}$  \\
 &  $\rm CC0\pi0p:15$-$20^\circ$   & $\pm27.4$ & $^{+10.8}_{-9.2}$ & $^{+4.8}_{-4.4}$ & $\pm9.0$ & $^{+31.2}_{-30.6}$  \\
 &  $\rm CC0\pi0p:20$-$25^\circ$   & $\pm23.8$ & $^{+10.8}_{-9.2}$ & $^{+16.2}_{-15.2}$ & $\pm7.1$ & $^{+31.6}_{-30.6}$  \\
 &  $\rm CC0\pi0p:25$-$30^\circ$   & $\pm25.7$ & $^{+10.4}_{-8.8}$ & $^{+7.9}_{-6.8}$ & $\pm9.0$ & $^{+30.2}_{-29.4}$  \\
 &  $\rm CC0\pi0p:30$-$180^\circ$  & $\pm29.3$ & $^{+10.4}_{-9.0}$ & $^{+29.1}_{-32.2}$ & $\pm11.1$ & $^{+44.0}_{-45.8}$  \\
\hline
 &  $\rm CC Total$                 & $\pm8.8$ & $^{+9.6}_{-8.1}$ & $^{+19.5}_{-17.5}$ & $\pm5.3$ & $^{+24.0}_{-21.8}$  \\
 &  $\rm CC0\pi0p:0$-$30^\circ$    & $\pm5.0$ & $^{+10.6}_{-9.1}$ & $^{+4.5}_{-4.3}$ & $\pm3.8$ & $^{+13.0}_{-11.8}$  \\
\hline
\hline
$\rm \sigma_{H_{2}O}/\sigma_{CH}$
 &  $\rm CCother$                  & $\pm49.9$ & $^{+0.6}_{-0.6}$ & $^{+107.8}_{-86.9}$ & $\pm51.7$ & $^{+129.6}_{-112.7}$  \\
 &  $\rm CC0\pi0p:0$-$5^\circ$     & $\pm36.6$ & $^{+1.7}_{-1.6}$ & $^{+45.2}_{-34.8}$ & $\pm23.6$ & $^{+62.8}_{-55.8}$  \\
 &  $\rm CC0\pi0p:5$-$10^\circ$    & $\pm51.5$ & $^{+1.5}_{-1.5}$ & $^{+17.5}_{-23.0}$ & $\pm15.4$ & $^{+56.6}_{-58.5}$  \\
 &  $\rm CC0\pi0p:10$-$15^\circ$   & $\pm31.6$ & $^{+1.1}_{-1.1}$ & $^{+32.6}_{-26.8}$ & $\pm17.1$ & $^{+48.5}_{-44.9}$  \\
 &  $\rm CC0\pi0p:15$-$20^\circ$   & $\pm40.3$ & $^{+0.8}_{-0.8}$ & $^{+19.7}_{-20.3}$ & $\pm16.8$ & $^{+47.9}_{-48.1}$  \\
 &  $\rm CC0\pi0p:20$-$25^\circ$   & $\pm40.0$ & $^{+0.9}_{-0.9}$ & $^{+17.2}_{-17.7}$ & $\pm17.0$ & $^{+46.8}_{-46.9}$  \\
 &  $\rm CC0\pi0p:25$-$30^\circ$   & $\pm52.3$ & $^{+1.6}_{-1.6}$ & $^{+10.0}_{-8.5}$ & $\pm17.6$ & $^{+56.1}_{-55.9}$  \\
 &  $\rm CC0\pi0p:30$-$180^\circ$  & $\pm37.0$ & $^{+1.4}_{-1.4}$ & $^{+36.4}_{-37.9}$ & $\pm26.0$ & $^{+58.0}_{-59.0}$  \\
\hline
 &  $\rm CC Total$                 & $\pm12.6$ & $^{+0.4}_{-0.4}$ & $^{+17.1}_{-19.0}$ & $\pm10.6$ & $^{+23.8}_{-25.1}$  \\
 &  $\rm CC0\pi0p:0$-$30^\circ$    & $\pm7.9$ & $^{+0.5}_{-0.5}$ & $^{+6.2}_{-5.8}$ & $\pm7.0$ & $^{+12.3}_{-12.1}$  \\
\hline
\hline
\end{tabular}
}
\end{table}
\begin{table}[!h]
\centering
\caption{Summary of systematic uncertainties on the \nmbm cross-section [\%].}
\label{tab:totalerror2}
\scalebox{0.9}{
\begin{tabular}{clccccc} 
\hline
\hline
 & True phase space  & Statistics & Neutrino & Neutrino     & Detector & Total \\
 &                   &            &   flux   & interactions & response &       \\
\hline
\hline
$\rm \sigma_{H_{2}O}$
 &  $\rm CCother$                  & $\pm13.6$ & $^{+8.3}_{-7.0}$ & $^{+14.3}_{-14.1}$ & $\pm17.8$ & $^{+27.8}_{-27.4}$  \\
 &  $\rm CC0\pi0p:0$-$5^\circ$     & $\pm29.3$ & $^{+10.4}_{-8.9}$ & $^{+25.9}_{-21.6}$ & $\pm19.2$ & $^{+44.8}_{-42.1}$  \\
 &  $\rm CC0\pi0p:5$-$10^\circ$    & $\pm39.4$ & $^{+11.0}_{-9.2}$ & $^{+8.2}_{-7.6}$ & $\pm13.8$ & $^{+44.0}_{-43.4}$  \\
 &  $\rm CC0\pi0p:10$-$15^\circ$   & $\pm16.8$ & $^{+10.9}_{-9.4}$ & $^{+21.5}_{-21.2}$ & $\pm13.4$ & $^{+32.3}_{-31.6}$  \\
 &  $\rm CC0\pi0p:15$-$20^\circ$   & $\pm32.0$ & $^{+10.7}_{-9.1}$ & $^{+18.0}_{-17.7}$ & $\pm14.1$ & $^{+40.7}_{-40.2}$  \\
 &  $\rm CC0\pi0p:20$-$25^\circ$   & $\pm19.1$ & $^{+10.6}_{-9.1}$ & $^{+12.9}_{-8.9}$ & $\pm18.2$ & $^{+31.2}_{-29.3}$  \\
 &  $\rm CC0\pi0p:25$-$30^\circ$   & $\pm29.4$ & $^{+10.7}_{-9.2}$ & $^{+12.2}_{-10.5}$ & $\pm11.9$ & $^{+35.6}_{-34.7}$  \\
 &  $\rm CC0\pi0p:30$-$180^\circ$  & $\pm31.3$ & $^{+10.3}_{-8.7}$ & $^{+30.7}_{-27.0}$ & $\pm23.3$ & $^{+50.7}_{-48.3}$  \\
\hline
 &  $\rm CC Total$                 & $\pm6.4$ & $^{+9.3}_{-8.0}$ & $^{+5.8}_{-5.6}$ & $\pm7.4$ & $^{+14.7}_{-13.8}$  \\
 &  $\rm CC0\pi0p:0$-$30^\circ$    & $\pm5.5$ & $^{+10.6}_{-9.1}$ & $^{+5.0}_{-4.0}$ & $\pm5.2$ & $^{+13.9}_{-12.5}$  \\
\hline
\hline
$\rm \sigma_{CH}$
 &  $\rm CCother$                  & $\pm7.1$ & $^{+8.6}_{-7.3}$ & $^{+39.5}_{-37.3}$ & $\pm9.6$ & $^{+42.1}_{-39.9}$  \\
 &  $\rm CC0\pi0p:0$-$5^\circ$     & $\pm19.4$ & $^{+10.4}_{-8.8}$ & $^{+13.0}_{-12.9}$ & $\pm6.4$ & $^{+26.3}_{-25.7}$  \\
 &  $\rm CC0\pi0p:5$-$10^\circ$    & $\pm25.7$ & $^{+10.8}_{-9.2}$ & $^{+6.5}_{-6.4}$ & $\pm5.1$ & $^{+29.0}_{-28.5}$  \\
 &  $\rm CC0\pi0p:10$-$15^\circ$   & $\pm19.0$ & $^{+10.0}_{-8.6}$ & $^{+4.8}_{-4.7}$ & $\pm6.1$ & $^{+22.8}_{-22.2}$  \\
 &  $\rm CC0\pi0p:15$-$20^\circ$   & $\pm22.8$ & $^{+10.7}_{-9.2}$ & $^{+4.2}_{-4.4}$ & $\pm8.4$ & $^{+26.9}_{-26.4}$  \\
 &  $\rm CC0\pi0p:20$-$25^\circ$   & $\pm18.2$ & $^{+10.3}_{-8.9}$ & $^{+7.7}_{-6.3}$ & $\pm6.9$ & $^{+23.3}_{-22.3}$  \\
 &  $\rm CC0\pi0p:25$-$30^\circ$   & $\pm21.2$ & $^{+9.9}_{-8.5}$ & $^{+8.5}_{-8.3}$ & $\pm8.4$ & $^{+26.3}_{-25.7}$  \\
 &  $\rm CC0\pi0p:30$-$180^\circ$  & $\pm23.5$ & $^{+10.3}_{-9.0}$ & $^{+24.5}_{-28.4}$ & $\pm11.3$ & $^{+37.3}_{-39.6}$  \\
\hline
 &  $\rm CC Total$                 & $\pm5.8$ & $^{+9.4}_{-7.9}$ & $^{+13.2}_{-12.0}$ & $\pm3.6$ & $^{+17.6}_{-15.9}$  \\
 &  $\rm CC0\pi0p:0$-$30^\circ$    & $\pm4.2$ & $^{+10.3}_{-8.9}$ & $^{+2.7}_{-2.9}$ & $\pm3.4$ & $^{+11.9}_{-10.8}$  \\
\hline
\hline
$\rm \sigma_{H_{2}O}/\sigma_{CH}$
 &  $\rm CCother$                  & $\pm16.0$ & $^{+0.6}_{-0.7}$ & $^{+49.4}_{-44.6}$ & $\pm20.6$ & $^{+55.9}_{-51.9}$  \\
 &  $\rm CC0\pi0p:0$-$5^\circ$     & $\pm33.7$ & $^{+1.6}_{-1.6}$ & $^{+27.6}_{-27.3}$ & $\pm20.9$ & $^{+48.3}_{-48.2}$  \\
 &  $\rm CC0\pi0p:5$-$10^\circ$    & $\pm41.5$ & $^{+1.6}_{-1.6}$ & $^{+9.5}_{-10.3}$ & $\pm15.2$ & $^{+45.2}_{-45.4}$  \\
 &  $\rm CC0\pi0p:10$-$15^\circ$   & $\pm23.6$ & $^{+1.7}_{-1.7}$ & $^{+24.6}_{-24.2}$ & $\pm15.2$ & $^{+37.4}_{-37.1}$  \\
 &  $\rm CC0\pi0p:15$-$20^\circ$   & $\pm37.8$ & $^{+1.5}_{-1.4}$ & $^{+13.9}_{-13.9}$ & $\pm17.3$ & $^{+43.8}_{-43.8}$  \\
 &  $\rm CC0\pi0p:20$-$25^\circ$   & $\pm26.3$ & $^{+0.9}_{-0.9}$ & $^{+11.1}_{-10.1}$ & $\pm20.3$ & $^{+35.1}_{-34.8}$  \\
 &  $\rm CC0\pi0p:25$-$30^\circ$   & $\pm34.6$ & $^{+1.7}_{-1.7}$ & $^{+15.3}_{-14.6}$ & $\pm15.0$ & $^{+40.8}_{-40.5}$  \\
 &  $\rm CC0\pi0p:30$-$180^\circ$  & $\pm41.1$ & $^{+1.9}_{-1.8}$ & $^{+40.2}_{-35.3}$ & $\pm27.4$ & $^{+63.8}_{-60.8}$  \\
\hline
 &  $\rm CC Total$                 & $\pm8.7$ & $^{+0.3}_{-0.3}$ & $^{+14.2}_{-15.4}$ & $\pm8.6$ & $^{+18.8}_{-20.0}$  \\
 &  $\rm CC0\pi0p:0$-$30^\circ$    & $\pm6.9$ & $^{+0.5}_{-0.6}$ & $^{+5.2}_{-4.3}$ & $\pm6.5$ & $^{+10.9}_{-10.5}$  \\
\hline
\hline
\end{tabular}

}
\end{table}

\clearpage

%%% References %%%
%\bibliographystyle{ptephy}
\bibliographystyle{wagasci_paper}
\bibliography{wagasci_paper}

\end{document}